\documentclass{mn2e}
\usepackage{graphics}
\usepackage{graphicx}

\newcommand{\X} {\mathbf{x}}
\newcommand{\V} {\mathbf{v}}

\newcommand{\fxv} {f(\X,\V)}
\newcommand{\ddd}[2]{\frac{\partial #1}{\partial #2}}

\newcommand{\ff} {b}

\title{Phase-space structures I: \\ 
A comparison of 6D density estimators
}

\author[M. Maciejewski, S. Colombi, C. Alard, F. Bouchet, C. Pichon]{M. Maciejewski$^{1,2}$, S. Colombi$^{1}$, C. Alard$^{1}$ , F. Bouchet$^{1}$, C. Pichon$^{1}$ \thanks{E-mail: maciejewski.michal@gmail.com (MM); colombi@iap.fr (SC); alard@iap.fr (CA); bouchet@iap.fr (FB);
pichon@iap.fr (CP) } \\$^{1}$Institut d'Astrophysique de Paris, CNRS UMR 7095 \& UPMC, 98 bis boulevard Arago, 75014 Paris, France 
\\$^{2}$Max-Planck-Institut f\"{u}r Astrophysik, Garching, Karl-Schwarzschild-Stra\ss e 1, 85741 Garching bei M\"{u}nchen, Germany}

\begin{document}


\maketitle

\begin{abstract}
In the framework of particle-based Vlasov systems, this paper reviews and analyses
different methods recently proposed in the literature to identify
neighbours in six dimensional space (6D) and estimate the
corresponding phase-space density.
Specifically, it compares Smooth Particle Hydrodynamics (SPH) methods
based on tree partitioning to 6D Delaunay tessellation. 
This comparison is carried out on statical and dynamical realisations of single halo profiles,
paying particular attention to the unknown scaling, $S_{\rm G}$, used to
relate the spatial dimensions to the velocity dimensions. 

It is found that, in practice, the methods with local adaptive metric
provide the best phase-space estimators. They make use of a Shannon
entropy criterion combined with a binary tree partitioning and
with subsequent SPH interpolation using 10 to 40 nearest neighbours.
We note that the local scaling $S_{\rm L}$ implemented by such methods, which
enforces local isotropy of the distribution function,
can vary  by about one order of magnitude in different
regions within the system. It presents a bimodal distribution, 
in which one component is dominated by the main part
of the halo and the other one is dominated by the substructures
of the halo.

While potentially better than SPH techniques, since
it yields an optimal estimate of the local softening volume
(and therefore the local number of neighbours required to perform the
interpolation), the Delaunay tessellation in fact generally 
poorly estimates the phase-space distribution function. 
Indeed, it requires, prior to its implementation, the choice of a global
scaling $S_{\rm G}$. We propose two simple 
but efficient methods to estimate $S_{\rm G}$ that yield a good 
global compromise. However, the Delaunay interpolation still remains
quite sensitive to local anisotropies in the distribution.

To emphasise the advantages of 6D analysis versus
traditional 3D analysis, 
we also compare realistic six dimensional phase-space density
estimation with the proxy proposed earlier in the literature,
$Q=\rho/\sigma^3$, where  $\rho$ is the local three dimensional (projected) density and
$3 \sigma^2$ is the local three dimensional velocity dispersion.
We show that $Q$ only corresponds to a rough approximation of the
true phase-space density, and is not able to capture all the details of the distribution in
phase-space, ignoring, in particular, filamentation and tidal streams.
\end{abstract}

\begin{keywords}
methods: data analysis, methods: numerical, galaxies: haloes, galaxies: structure, cosmology: dark matter
\end{keywords}

\section{Introduction}
There are many methods to analyse dark matter haloes structures. A standard
approach involves investigating spherically averaged density profiles, 
such as  the Hernquist profile \cite{Hernquist1990},
the NFW profile \cite{NFW97}, the Moore profile 
\cite{Moore1998,Moore1999} and the Stoehr profile \cite{Stoehr2006}. 
More sophisticated  methods
developed recently  involve different elliptical density 
profiles \cite{Jing2002,Hayashi2007}. An other alternative
consists of analysing velocity profiles, e.g., Romano-Diaz \&
van de Weygaert (2007), for a review.

Other investigations look in more details at halo detection
as well as their internal substructures, the subhaloes. They 
usually use a two steps procedure: they first find 
haloes and substructures in position space, then use
velocity information to apply binding
criteria. Many such schemes are found in the literature, 
the simplest being the Friend-Of-Friend (FOF)
algorithm \cite{FOF}. More advanced methods rely for instance 
on the SKID algorithm \cite{Stadel} or on 
SUBFIND \cite{Springel2001}.

However, thanks to increased computational power, it now  becomes
possible to perform more detailed analyses that combine
simultaneously velocity and position information. Indeed, modern
simulations now reach  enough resolution to identify
structures and substructures in the full, six dimensional phase-space.
Recent investigations in this topic studied phase-space dark matter
profiles \cite{Taylor2001}, phase-space density estimation by using six dimensional
Delaunay Tessellation (SHESHDEL) \cite{Arad2004}, or by using binary tree methods with smoothing
(FiEstAS) \cite{Binney2004}, and a variety of different binary tree and six dimensional SPH
methods with local adaptive metric in the EnBiD package \cite{Sharma2006}.

Two noticeable results were derived  within these investigations:
the measurement of a universal
logarithmic slope for the phase-space density $f$ as a function of radius $r$,
$f(r)\sim r^{-\alpha}$, with $\alpha\sim1.875$ \cite{Taylor2001}, and the
observation of a universal profile for the phase-space volume occupation
function, $v(f)\propto f^{-2.5\pm0.05}$ \cite{Arad2004,Binney2004}.

These results depend on the quality of the phase-space density estimators, 
a topic to which we devote this paper. We carefully analyse
and cross compare the SHESHDEL, FiEstAS and EnBiD estimators. 

This paper is organised as follows. Section~\ref{sec:est} describes
the various generic\footnote{i.e. applicable to systems without 
specific symmetries.} phase-space estimators and the corresponding concepts.
We pay particular attention to the issue of the unknown scaling, $S_{\rm G}$,
which  relates position coordinates and velocity coordinates
prior to the phase-space distribution function measurement. 
In Section~\ref{sec:numexp} we test the phase-space estimators
in three realisations of a halo profile: (i) a pure Hernquist isotropic
halo, (ii) a composite Hernquist halo (a main Hernquist
component with Hernquist subhaloes) and
(iii) a halo extracted from a standard Cold Dark Matter (CDM)
cosmological simulation. To have a better understanding of the results,
more thorough analyses are performed in
Section~\ref{sec:additiona}  focusing on  (i) the local number of
neighbours built by the Delaunay tessellation and on (ii) the local scaling
between positions and velocities given by the adaptive metric
of EnBiD. Section~\ref{sec:secQ} shows the advantages of full phase-space analysis,
with respect to more classical approaches such as the  proxy $Q=\rho/\sigma^3$.
Finally, Section~\ref{sec:conc} wraps up.

\section{Phase-space density estimation}
\label{sec:est}

There are a few common approaches to measure six 
dimensional phase-space density, $\fxv$, for unrelaxed systems. 

A straightforward method involves dividing  phase-space into a Cartesian
grid and approximating phase-space density by counting particles in
each bin. 
While this clearly works quite well in three-dimensional space, 
it starts to be problematic in six dimensions. 
Even if we choose a poor quality resolution, e.g. $100$ bins along
each axis, we get in the end a very large number of cells,
e.g. $10^{12}$, and, for modern simulations with e.g. $10^{7}$
particles, almost all the cells will be empty. 
This basic example shows that for improved phase-space
estimation, one needs to go well beyond the
naive binning algorithm. Notice as well that to achieve a level of
detail in phase-space comparable to what is usually obtained in position
space, one needs a simulation with an extremely large number of
particles.

A more sophisticated, frequently used method for density estimation
in position space,  uses smoothing with $k$ nearest neighbours found with standard
tree techniques; it can be easily 
generalised to the six dimensional case. Assuming for the sake
of simplicity that all particles have the same mass $m_p$, if, for
each particle, $k$ neighbours are enclosed in a six dimensional ball
of volume $V_6$, then the local phase-space density can for instance be measured with
the simple following estimator, $m_p k/V_6$, which corresponds to
a top hat kernel. In practice, more sophisticated kernels are used, 
i.e. each neighbour contributes to the measured
density with a weight defined by a smooth function, 
usually a Smooth Particle Hydrodynamics (SPH) kernel. 
This kind of algorithm was proposed for phase-space density 
estimation by Sharma and Steinmetz (2006) (hereafter S06). It
however requires the proper set up of a metric in six-dimensional
space (velocity/position scaling). 

In this paper we investigate more accurate
algorithms developed recently in the literature, including
improvements of the above SPH technique.

The first method, discussed in \S~\ref{sec:DTFE}, 
relies on six dimensional Delaunay tessellation (Arad et al., 2004; hereafter Arad04). 
The big advantage of this method is that it is parameter free, fully
adaptive, while each particle has a natural neighbourhood. 
In practice, however, the Delaunay tessellation needs some additional smoothing.
It is also very time and memory consuming \cite{Arad2004,Weygaert2007}.
It requires, similarly as the straightforward SPH method just
mentioned above, a proper set up of a metric in six-dimensional space.

The second group of algorithms was proposed by Ascasibar \& Binney
(2004, hereafter A04) and improved by S06. 
The first step of their method, detailed in \S~\ref{sec:treepart}, 
is simple and robust. Space is divided
with the help of a binary tree into disjoint 
hyperboxes with one particle in each leaf node. 
Since each particle is in one hyperbox with volume $V$, its local
phase-space density could be directly estimated from 
the equation $f=m_p/V$. Yet the phase-space density  derived from this
estimator is quite noisy: it is almost impossible to use it for practical
purposes. 
Hence additional steps were proposed to make it useful. 
First, the binary tree may be improved with the help of a Shannon 
entropy criterion combined with boundary particles correction
(S06). 
Secondly, some additional smoothing should be performed.
There are few options to do so, as proposed by A04 and S06 and
described in \S~\ref{sec:mysmoothing},  ranging
from (a) a hyperbox smoothing following the philosophy of the SPH method, 
(b) a SPH method with a local adaptive metric, to (c)
anisotropic SPH methods. The main advantage of this type of algorithm
compared to the tessellation methods is time and memory consumption.

In the next sections, we describe each of these methods in turn and follow
with a detailed discussion on the issue of
position/velocity scaling (\S~\ref{sec:posvelcor}). 
The reader may refer to Table~\ref{table:algorithms} and to the
summary in \S~\ref{sec:conc}, if needed.
\begin{table*}
\caption[]{The various 6D estimators tested in this paper}
\begin{tabular}{|l|l|} \hline
The estimator name & Description \\ \hline
SPH & Smoothing with spherical Epanechikov kernel using $N$ neighbours, global scaling \\ \hline
SPH-AM & Smoothing with spherical Epanechikov kernel using $N$ neighbours, local adaptive metric \\ \hline
ASPH-AM & Anisotropic smoothing with ellipsoidal Epanechikov kernel, using $N$ neighbours, local adaptive metric \\ \hline
FiEstAS & Smoothing with the hyperbox kernel, local adaptive metric \\ \hline
DTFE & Estimation from the Delaunay tessellation (equation~\ref{eq:DTFEsimple}), global scaling \\ \hline
smooth DTFE & Estimation from the Delaunay tessellation with spherical smoothing (equation~\ref{eq:DTFEsph}),
global scaling \\ \hline
\label{table:algorithms}
\end{tabular} 
\end{table*}

\subsection{The Delaunay Tessellation}
\label{sec:DTFE}
The idea of using a Delaunay tessellation was primarily implemented
to estimate density and velocity fields in cosmological simulations
\cite{Bernardeau,2003sca..book..483S}. The corresponding algorithm,
called the  Delaunay Tessellation Field Estimator (DTFE),  was also
used to demonstrate the advantages of the tessellation over SPH methods
\cite{Pelupessy2003}. In particular, the DTFE method
 better captures the abrupt transitions
between regions of different densities and
provides a better estimate for high densities.

The main parts of the DTFE algorithm were used by Arad04 to develop their
six dimensional phase-space density estimator called 
SHESHDEL. Beside the position-velocity scaling problem that is addressed in
\S~\ref{sec:posvelcor} and in \S~\ref{sec:posvelcor2},
the method itself is parameter free and presents well behaved
statistical properties. Its main disadvantage is that it is
computationally costly, although it scales
like $N^{1.1}\log N$ (Arad, in preparation), where $N$ is the number
of particles: the construction of a Delaunay
tessellation of approximately $N\sim10^{6}$ particles requires almost three days of
calculation on a modern computer and the full output of $10^9$ Delaunay cells
amount to $40$ GB of data in the end.

From a 6D Delaunay tessellation, it is easy to 
estimate the phase-space density, $\fxv$. 
Space is indeed partitioned into joint but non
overlapping 6-dimensional polyhedrons - Delaunay
cells, each one defined by 7 vertices. There
is an unique 6  dimensional sphere passing through these
7 vertices, which by definition of the tessellation, 
does not encompass any other particle from the sample. 
Let $\{D_i^1,D_i^2,...,D_i^{N_i}\}$ be the $N_i$ Delaunay 
cells around particle $i$. We can define a macro \textit{Voronoi} 
cell $W_i$ by joining all Delaunay cells containing particle $i$:
\begin{equation}
W_i=\bigcup_{j=1,...,N_i} D_i^j.
\end{equation}
Then it is straightforward to define an estimate
of the phase-space density for each particle $i$ of mass 
$m_p$ as:
\begin{equation} \label{f_i}
f_i=7\frac{m_p}{\left\vert W_i \right\vert},
\label{eq:DTFEsimple}
\end{equation}
where $|W_i|$ is the volume of the macro cell and the factor $7$  
 accounts for  the fact that each Delaunay cell contributes
to the density of 7 particles.
In practice, as mentioned earlier, the corresponding estimated phase-space density  is very
noisy, and one must introduce some additional smoothing.
Let 
\begin{equation}
f_{D_i}=\frac{1}{7}\sum_{j\in{D_i}}f_j\,,
\end{equation}
be the average phase-space density defined for Delaunay cell $D_i$.
One can then define a smoother phase-space density estimator as:
\begin{equation}
f''_i=\frac{\sum\limits_{j=1,...,N_i}f_{D_i^j}|D_i^j|}{\left\vert W_i \right\vert},
\end{equation}
where $j$ indexes all Delaunay cells around particle $i$ and $|D_i^j|$
represents the volume of each Delaunay cell.
\begin{figure}
\includegraphics[width=8cm,height=8cm]{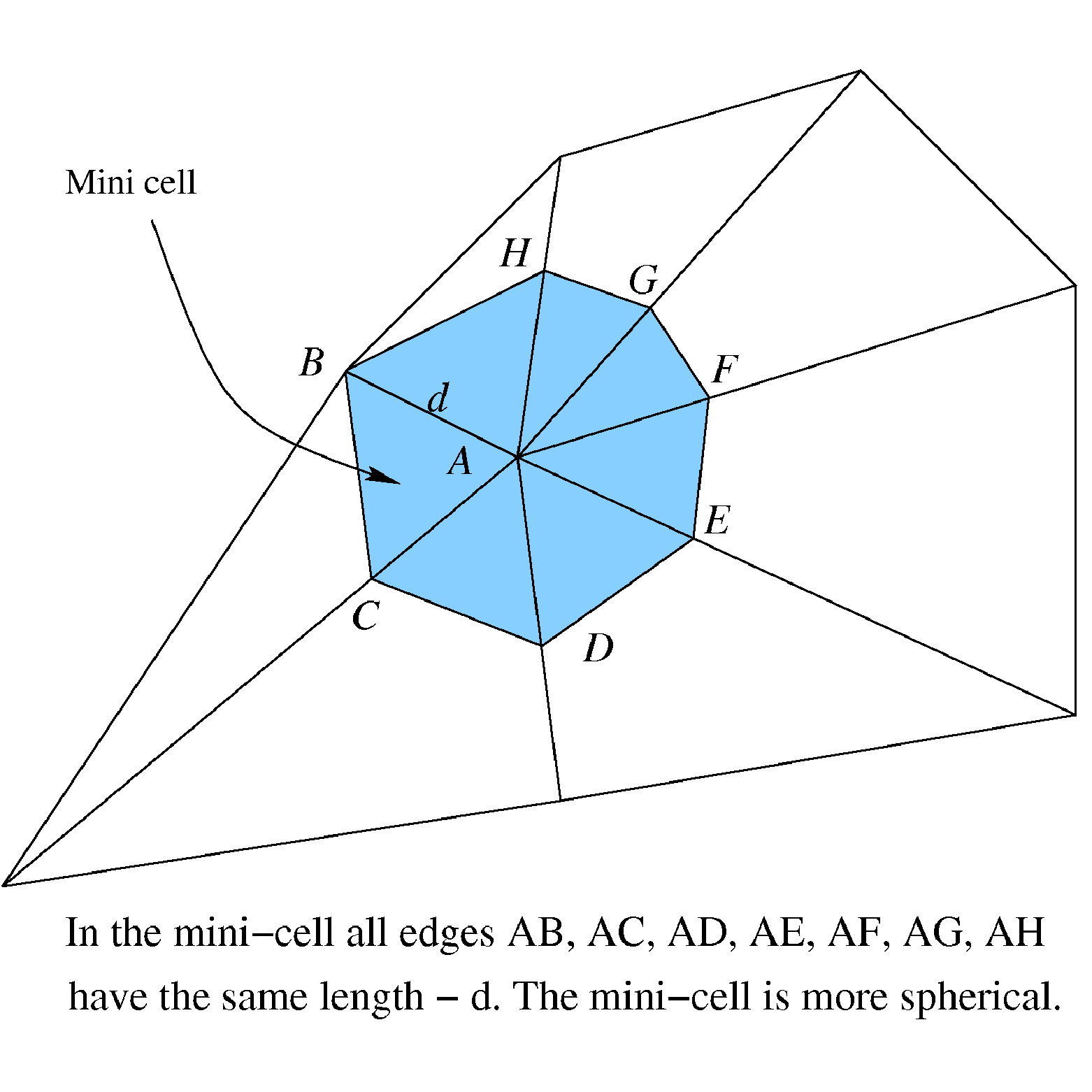}
\caption{A method of smoothing which is appropriate for a Delaunay
tessellation and corrects for local anisotropies: 
it involves redefining the local volume 
in such a way that one gets in the end a more spherical mini-cell.}
\label{minicell_fig}
\end{figure}

For simulations without e.g. periodic boundaries, the  phase-space density 
of particles near the edge of the computing domain 
can be underestimated. This is for instance the case when
the sample has been cut from a bigger cosmological volume.
To cope with this edge effect problem, 
one can introduce another definition for
the smooth phase-space density estimator. 
Consider particle $i$, surrounded by its Delaunay cells $D_j$,  
and let $d$ be the distance between this particle
and its closest neighbour (Figure~\ref{minicell_fig}). 
Then, the mini-cell around particle $i$ is defined as the collection of
Delaunay cells $D'_j$, which are similar to $D_j$ but are
scaled in such a way that each edge in $D'_j$ containing particle $i$ 
is exactly of length $d$. The volume of $D'_j$ reads
\begin{equation} 
|D'_i|=|D_i|\prod_{j\in{D_i}}\frac{d^6}{|e_{ij}|},
\end{equation}
where $|e_{ij}|$ is the length of the segment joining particle $i$ and $j$ in 
Delaunay cell $D_i$. As a result, a natural phase-space density
estimator reads
\begin{equation}
f'''_i=\frac{\sum\limits_{j=1,...,N_i}f_{D_i^j}|{D'}_i^j|}{\sum\limits_{j=1,...,N_i}|{D'}_i^j|}.
\end{equation}
However, it might be better to use linear interpolation to estimate the
phase-space density in each mini-cell, to perform additional local
noise filtering:
\begin{equation} 
f_{D'_i}=f_{D_i}+\frac{d}{7}\sum_{j\in{D_i}}{\frac{f_i-f_j}{|e_{ij}|}}.
\end{equation}
In the end we can thus define a phase-space density estimator 
with interpolated density as follow:
\begin{equation}
f'_i=\frac{\sum\limits_{j=1,...,N_i}f_{{D'}_i^j}|{D'}_i^j|}{\sum\limits_{j=1,...,N_i}|{D'}_i^j|}.
\label{eq:DTFEsph}
\end{equation}
The mini-cells are more regular and spherical, so  both  
phase-space density estimators $f'$ and
$f'''$ are expected to be less sensitive to local fluctuations and 
local anisotropies due to noise and edge effects.

Note that all of the above smoothing methods are based on
Natural Neighbour Interpolation techniques \cite{Weygaert2007}.
In what shall follow, although we studied all the estimators,
$f_i$, $f''_i$, $f'_i$ and $f'''_i$, we shall present explicit
results only on $f'_i$ (equation~\ref{eq:DTFEsph}) and on $f_i$
(equation~\ref{eq:DTFEsimple}).

\subsection{Tree partitioning}
\label{sec:treepart}
As discussed in the introduction of \S~\ref{sec:est}, the main point of the 
algorithm FiEstAS~\footnote{Field Estimator for Arbitrary Spaces} 
proposed by A04 and later improved
by S04 in the EnBiD~\footnote{Entropy based Binary
Decomposition} implementation, is the division of space into a binary tree. 
In FiEstAS, the splitting axis is chosen alternatively 
between position space and velocity space, then in each of these
respective subspaces, the axis with highest elongation, 
$\langle x_i^2 \rangle- \langle x_i\rangle^2$, is split. 
This splitting criterion helps the cells to
preserve a shape as cubic as possible. 

However, a visual inspection of position and
velocity diagrams of typical simulations (Figure~\ref{figure2bb}) shows
that position space contains more structures, and thus
more information than velocity space. As a result, one can argue
that for optimal accuracy, the splitting should occur more often in position space
than in velocity space. This observation was used in the EnBiD algorithm 
to define a better splitting criterion. Before splitting occurs, one has
to find the subspace (velocity or position) in which it should 
be performed. To do that, the Shannon entropy, $S$, is calculated after
dividing each subspace into $N$ equal size bins:\footnote{the choice
of S06 is $N$ to be equal to the number of particles contained in the subspace.}
\begin{equation}
 S = - \sum_{i=1}^{N}\frac{n_i}{N}\log{\frac{n_i}{N}},
\end{equation}
where $n_i$ is the number of particles in each bin. The subspace
which has to be splitted is the one with smallest entropy. Finally, the
direction of splitting is chosen again using the highest elongation
criterion, to preserve a close to cubic shape.

As for Delaunay tessellation, correction for edge
effects is crucial in the binary tree partition algorithm.
To illustrate that point, it was shown in S06 
that for $10^6$ particles uniformly distributed in a 6D spherical region,
about $79\%$ of them lie near the border, compared to $5\%$ in the 3D
case. {This reflects the so called curse of dimensions}.
 The natural shape of local border in
the binary tree partition algorithm is an hypercube. When the data do
not preserve locally this shape, 
the volume occupied by the boundary particles tends to be overestimated, 
hence their phase-space density tends to be underestimated. This
bias is moreover expected to worsen and to propagate further away from the
edges if additional smoothing is performed.

Both FiEstAS and EnBiD redefine borders to correct for edge effects. 
While FiEstAS does it only for the tree leaves, EnBiD applies the 
correction to all the nodes of the tree, in order to
insure proper entropy calculation and to better 
estimate the phase-space density of small structures found in the 
halo.\footnote{See section 2.3 of S06 for more details.}

\subsection{Smoothing}
\label{sec:mysmoothing}

\begin{figure}
\includegraphics[width=8cm,height=5.87cm]{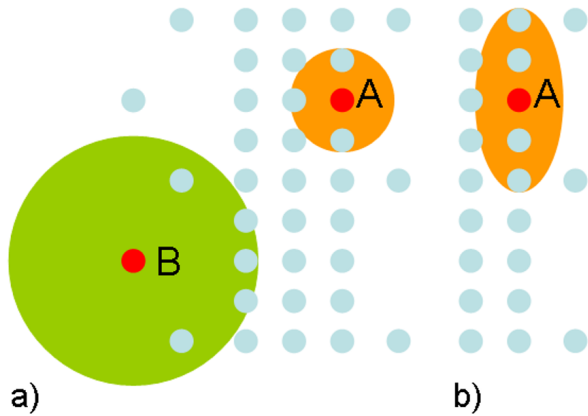}
\caption{Biases in the SPH density estimation and potential advantages of
  anisotropic SPH methods: a) SPH method: 
density is underestimated for particle A and overestimated for
  particle B; b) ASPH method better traces local structures and
  should lead to better density estimation.}
\label{sph_fig}
\end{figure}

From these tree methods, one could estimate naively 
the phase-space density by exploiting directly the information
stored in the tree structure, as argued in the end of the introduction
of \S~\ref{sec:est}, but measurements performed that way would
be rather noisy: additional interpolation, 
should be applied to the data in order to achieve a good measurement of
phase-space density. A04 and S06 investigated a few  
smoothing procedures, that we discuss now.

Let us first describe the smoothing method proposed by A04 
(called later FiEstAS smoothing). The main idea comes from SPH techniques, 
but the smoothing kernel is a hyperbox rather than a hypersphere. 
This treatment avoids the need for a definition of a local metric. 
First, the mass of each particle is distributed uniformly over its 
leaf volume. Then, a hyperbox of volume $V_s$, centred on this leaf
and with the same axis ratio, is found,
such that it contains a mass $M_s$, which basically defines the kernel
size. Local phase-space density is then  calculated from the
equation $f=M_s/V_s$. In our investigations, we shall use mainly the
$M_s=2 m_p$ value proposed by S06 (while A04 suggest $M_s=10m_p$).

The other approach uses a classic SPH technique. For our investigations, 
we shall use the Epanechikov kernel
\begin{equation}
 W(x,h)=f_d\left\{ \begin{array}{cc}
 \displaystyle
  1-\sum_{i=1}6\left(\frac{x_i}{h_i}\right)^2\,, & 0\leq{x_i}/{h_i} \leq 1 \,,\\ 0 \,,& {x_i}/{h_i} > 1\,,\end{array} \right.
\end{equation}
with an additional bias correction. As mentioned in S06,
this estimator seems to give the best results for SPH phase-space estimation.

One of the disadvantages of the SPH method is the way it handles
strong transitions between regions of very different densities,
as illustrated by Fig~\ref{sph_fig}a. The key
point is that the SPH method is not able to capture correctly strong
variations of the density local curvature. 
Because of the spherical shape of the kernel
and the fixed number of neighbours used to perform the calculations, 
the density will be underestimated near the edge
of the regions with higher density (particle A), 
or more generally in regions with significant negative local
curvature. On the other hand the density will be overestimated 
near the edge of the low density
regions (particle B), or in regions with significant positive local curvature. 
One way of resolving this issue, and of better
capturing filamentary structures such as in Fig~\ref{sph_fig}a,
is to use an anisotropic SPH kernel, which adapts locally to the 
shape of structures and
is thus more appropriate to capture to local curvature variations. 
First, the $N_A$ nearest neighbours are found for each particle,
which allows one to compute a deformation tensor $H$, that is used to
define a local ellipsoid with $N_B$ nearest neighbours (usually
$N_A>N_B$).  Each particle contained in this ellipsoid
contributes to the phase-space density with the weights given by the
SPH kernel, scaled properly to take into account the ellipsoid shape
(Fig~\ref{sph_fig}b).

Note that, in addition to the issues described in Figure~\ref{sph_fig}, the 
density calculated with SPH methods presents some non trivial biases
due to local Poisson shot noise which can in part be corrected for
(see S06 for details).

\begin{figure*}
\includegraphics[width=17cm,height=9.89cm]{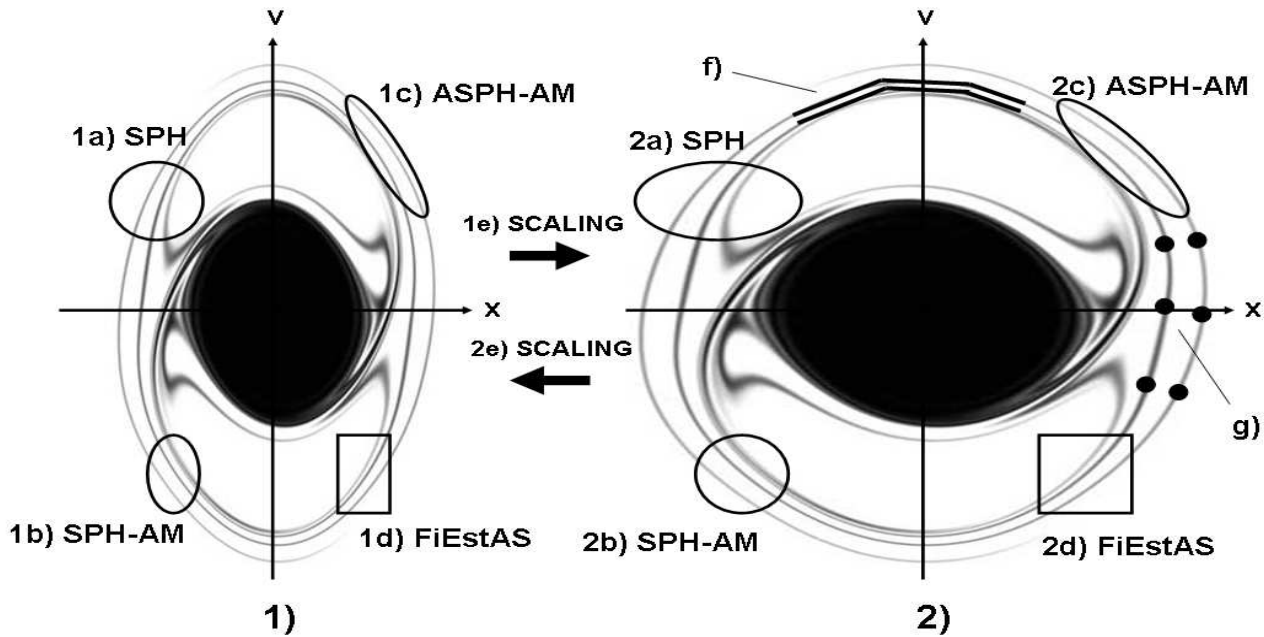}
\caption{Illustration from a 2D simulation
of Alard \& Colombi (2005)  of the  velocity-position scaling and the various
methods used to measure the coarse grained phase-space distribution
function. 1) The left panel shows the phase-space distribution
function in the ``global'' coordinate system used by the authors. 2) The right
panel shows a scaled version of it in such a way that the same
spread is observed in x and v coordinates. We call that a ``local''
system of coordinates as the scaling can be global as shown here, or
local, as discussed in the main text.
1a) SPH spherical kernel
in global coordinates,  2a) becomes
ellipsoidal in local coordinates and is not optimal; 
1b) SPH with a local Adaptive Metric presents an ellipsoidal kernel in global 
coordinates, 2b) but is set up in such a way that the kernel is
spherical in local coordinates;  1c) Anisotropic SPH with a local
Adaptive Metric in global coordinates, 2c) in local coordinates.
1d) FiEstAS smoothing in global coordinates 2d) presents a
hypercubical kernel in local coordinates;
f) local clouds resolve accurately phase-space structures, but g) particles
from cosmological simulations only sparse sample them.}
\label{sph_fig1}
\end{figure*}

\subsection{Position-velocity ``metric" correction}
\label{sec:posvelcor}
There is a last one important problem that arises when one aims to estimate
phase-space density. In order to perform
some local phase-space density estimation, one needs to find a way
to join position space and velocity space, or in other words to define a proper {\sl metric}
that allows one to estimate local distances and local volumes in
phase-space. This generic  problem  does not have
an obvious solution. Let us first discuss that issue and then detail
the implementations to resolve it for the
phase-space estimators used in this paper.
 
\subsubsection{Coarse grained versus fine grained phase-space density}
\label{sec:discuscaling}

The dynamics of dark-matter is usually modeled by a self-gravitational 
collisionless fluid which follows the so called Vlasov\footnote{also referred to as collisionless Boltzmann.}-Poisson equations:
\begin{equation}
\frac{\mathrm{d}f}{\mathrm{d}t}\equiv\ddd{f}{t}+\V\ddd{f}{\X}-\ddd{\phi}{\X}\ddd{f}{\V}=0,
\end{equation}
\begin{equation}
\nabla^2\phi(\X,t)=4\pi G \int{f(\X,\V,t) \mathrm{d}\V}.
\end{equation}
Because it is very difficult to solve these equations directly,
the continuous fluid formulation is usually approximated by collisionless
particles which follow the  classical gravitational Newtonian
equations, hence producing Monte Carlo 
realisations of this set,
with additional softening to maintain the forces bounded.
The most important question here is how this approximation affects
the phase-space density properties. Liouville theorem states that the phase-space 
distribution function remains constant along trajectories of the system
\begin{equation}
f(\X(t),\V(t),t)={\rm constant}.
\end{equation}
This is true for the smooth, \textit{fine-grained} phase-space density
$f$. In $N$-body simulations, it is in practice possible to probe
only the \textit{coarse-grained} phase-space density, $\bar{f}$, which is
the average of $f$ over a small but finite volume \cite{GD}. 
This quantity does not follow the  Liouville theorem anymore 
because of mixing processes occurring at small
scales \cite{Tremaine86,Arad2004}. Furthermore,
the measurement of ${\bar f}$ depends highly on the way the coarse
graining volume is defined, hence in particular on the local scaling to be
applied to velocities versus positions. 
To have a proper
measurement of ${\bar f}$ that approaches as much as possible
the fine grained distribution function from the dynamical point of view, or some sensible local
average of it, one would
need the knowledge of the whole dynamical history of the system.

One way to overcome this problem is to solve numerically Vlasov's
equations using a more sophisticated approach  than the simple
$N$-body method, where the phase-space
distribution function is modeled by small elements of metric,
such as ellipsoid ``clouds'', that sum up to a dynamically
meaningful coarse grained version of the distribution
function. This method is
discussed in detail and applied in 1+1 D in Alard \& Colombi (2005).
Of course the generalisation of such a method to 6D is
quite costly. However, a simple alternative, in the spirit of
this method, would be to attach to each particle of a $N$-body simulation the information
corresponding to the local phase-space volume (or the local metric), that would be
followed during the evolution
of the system \cite{vog2008}. 
Note that we  then follow 
a sparse sampling of the {\it fine grained} distribution function as
long as the dynamical effects due to the discrete particle
representation are negligible.
Then the appropriate shape for the phase-space element
used to measure the coarse-grained distribution function would
be given by a local average on a number of neighbouring particles in
phase-space, as  is
achieved  by the adaptive SPH method.

 Finally, if such an information is
not available, and without supplementary prior on the dynamical
state of the system, one can just try to find the 
best coordinate transform that preserves local isotropy within the coarse 
grained volume. This method basically assumes that the
systems evolved from a smooth distribution function. In that sense,
for cold dark matter haloes, it only traces correctly 
the coarse distribution after relaxation to a quasi-steady state 
(i.e. a few dynamical times after collapse). Note that a simple
application of this idea to find a global scaling between
position and velocities basically produces the system of coordinates
where the velocity scatter is of the same order of the positions scatter.

To illustrate this discussion, figure \ref{sph_fig1}.1
shows one of the outputs of a 2D phase-space simulation
of  Alard \& Colombi (2005), using the cloudy method (briefly
sketched above).  The system was evolved during approximately 40 
dynamical times from an initial top-hat distribution 
function slightly apodized at the edges.
Figure \ref{sph_fig1}.2 shows the same realisation, but the position 
coordinate is scaled in such a way that both velocities and positions 
show the same spread: this is the natural system of coordinate
for a global definition of the scaling to be used
between positions and velocities prior to the definition
of a small round coarse graining volume.

In the cold dark matter scenario, initial conditions
can by approximated by a three dimensional sheet 
(small dispersion in velocity space) immersed in six dimensional 
phase-space, that subsequently evolves in time and gains a complex shape
(without loosing its connectivity or volume as stated by the Liouville theorem). 
The equivalent of such a sheet in our 2D phase-space representation would
be e.g. the curve (f), accurately followed by many cloud elements.
As mentioned above, Vlasov-Poisson equations are usually numerically resolved
relying on a particle representation. After many time steps,
because of variations of the local force field, particles initially
close by depart from each other (g). Mixing processes occur at small scales,
and the phase-space sheet, poorly modeled by the particles, loses
its fine structures. The way the coarse-grained phase-space density 
$\bar{f}$ is calculated is 
shown on e.g. (2d). The use of a finite local volume for computing
$\bar{f}$ results in the averaging of the phase-space density
over many curves -- or sheets in six dimensions. 
From now on, unless otherwise stated, we shall skip the bar and
use the $f$ symbol for the coarse-grained phase-space density. 

\subsubsection{Solutions for the position-velocity scaling}
\label{sec:solsca}

Inspired by the discussion in previous paragraph, we now propose
two ways of fixing the position-velocity scaling:
\begin{itemize}
\item A {\em global} scaling factor between positions and velocities, that
will be applied to  the standard SPH and Delaunay Tessellation
algorithms. This global factor tries to make the phase-space
distribution function globally isotropic, i.e. with the same spread in velocities
and in positions.
\item A {\em local} scaling factor that depends on phase-space 
coordinate $(x,v)$, and that tries to make the phase-space
distribution function locally isotropic. This local scaling
factor is implemented by construction in FiEstAS and its
improvement, EnBiD,
as detailed below. It will be used as well when additional smoothing
is performed with SPH or ASPH  techniques. In that case we shall
denote the methods by SPH-AM and ASPH-AM, respectively.
\end{itemize}

\noindent {\it (i) Global scaling:}
\vskip 0.2cm

\noindent For the global scaling method depending on one parameter,
$S_{\rm G}$, the metric transform can be written in the form:
\begin{equation}
\left(\begin{array}{c} dx' \\ dv' \end{array}\right)
= \left(\begin{array}{cc}{1}/{\sqrt{S_{\rm G}}} & 0 \\ 0 & \sqrt{S_{\rm G}} \end{array}\right)
\left(\begin{array}{c} dx \\ dv \end{array}\right).
\end{equation}
We test here two different ways of setting $S_{\rm G}$, that apply to single
dynamical systems, such as the cold dark matter haloes analysed in this
paper. Such a global scaling finds a transformation that changes for
instance Fig. \ref{sph_fig1}.1 in
Fig. \ref{sph_fig1}.2.

The first scaling uses simple dynamical arguments that
lead to a comparable scatter in velocity and position space of
the phase-space distribution function, or in other
worlds to a ``more round'' shape of the cloud representing $f(x,v)$. 
It relies on the
fact that dark matter haloes are known to be well fitted by the NFW
profile \cite{NFW97}.  In particular, within that model, we
use  the relation
\begin{equation} \label{EQ10}
r_{200} = \frac{v_{200}}{10H(z)},
\end{equation}
where $r_{200}$ and $v_{200}$ are the viral radius and the viral
velocity of the halo, respectively, and 
$H(z)$ is the Hubble constant in units of $\mbox{km}~\mbox{s}^{-1}~\mbox{kpc}^{-1}$
\cite{Schneider2006}. 
For the haloes analysed in this paper,  $z=0$  and
$H(0)=0.1h$.  In that case, the natural set of coordinates,
that fixes properly the global scaling between positions and
velocities, simply uses distances expressed  in  
units of kpc~$h^{-1}$ and velocities in km~s$^{-1}$. In this
case the global scaling is set to $S_I=1.0h\mbox{ km}~\mbox{s}^{-1}~\mbox{kpc}^{-1}$.

The second method, more sophisticated, should reach approximately
the same scaling. It involves attempting  to enforce
global isotropy of the distribution of  particles in phase-space.
To achieve that, the distances of each particle to its closest
neighbour in position subspace and in velocity subspace,
are computed, which allows us to estimate 
the probability distribution functions of these distances,
$p(r)$ and $p(v)$. The global scaling $S_{\rm dist}$ between the two subspaces
is the one where the maximum of $p(r)$ and the maximum of $p(v)$
coincide. 

\vskip 0.2cm
\noindent {\it (ii) Local scaling:}
\vskip 0.2cm

\noindent Obviously, although it presents the advantage of being simple to
implement  and robust, the global scaling is not optimal. However, it
is possible to enforce, to some extent, local isotropy of the
phase-space distribution function by examining the
local neighbourhood of each particle, 
as  is implemented in the FiEstAS and EnBiD algorithms.

In FiEstAS the natural local scaling $S_{\rm L}$
between velocity subspace and position subspace is
simply determined from the axis ratio of the tree leaf 
containing the particle (which is equivalent to 
pass from Figure~\ref{sph_fig1}.1d to Figure~\ref{sph_fig1}.2d).
By construction of the tessellating
tree, the local isotropy between both subspaces should be preserved in the first
approximation, as the calculation of the final
smoothing hyperbox preserves this axis ratio. For the modification
of FiEstAS proposed by the  EnBiD algorithm, the splitting
of the binary tree is improved by the calculation of Shannon entropy,
which in principle warrants a  better local 
assignment of the metric frame.\footnote{Note that  the border corrections mentioned at the
  end of \S~\ref{sec:treepart} may have some significant impact on the local metric.}

Finally one can perform additional SPH or Anisotropic SPH interpolation in the
local metric frame determined by EnBiD, that lead to our SPH Adaptive Metric algorithms
(SPH-AM and ASPH-AM). Prior to SPH or ASPH interpolation, the
phase-space coordinates are scaled in such a  way that the
local hyperbox corresponding to the leaf containing the particle
becomes hypercubical (Figure~\ref{sph_fig1}.1d
to Figure~\ref{sph_fig1}.2d, to obtain Figure~\ref{sph_fig1}.2b and
Figure~\ref{sph_fig1}.2c). Of course the biases expected in SPH and ASPH
interpolations mentioned in the end of \S~\ref{sec:mysmoothing} 
are still present, even with this local metric approach.

To summarise, we see that FiEstAS method with EnBiD improvement 
corresponds to hyperbox smoothing with adaptive metric. The only
thing that changes between SPH-AM or ASPH-AM is the shape
of the kernel used to perform the smoothing. 

\section{Numerical experiments}
\label{sec:numexp}
\subsection{Hernquist profile}
\label{sec:hernquist}
\begin{figure}
\includegraphics[width=8.25cm,height=16.0cm]{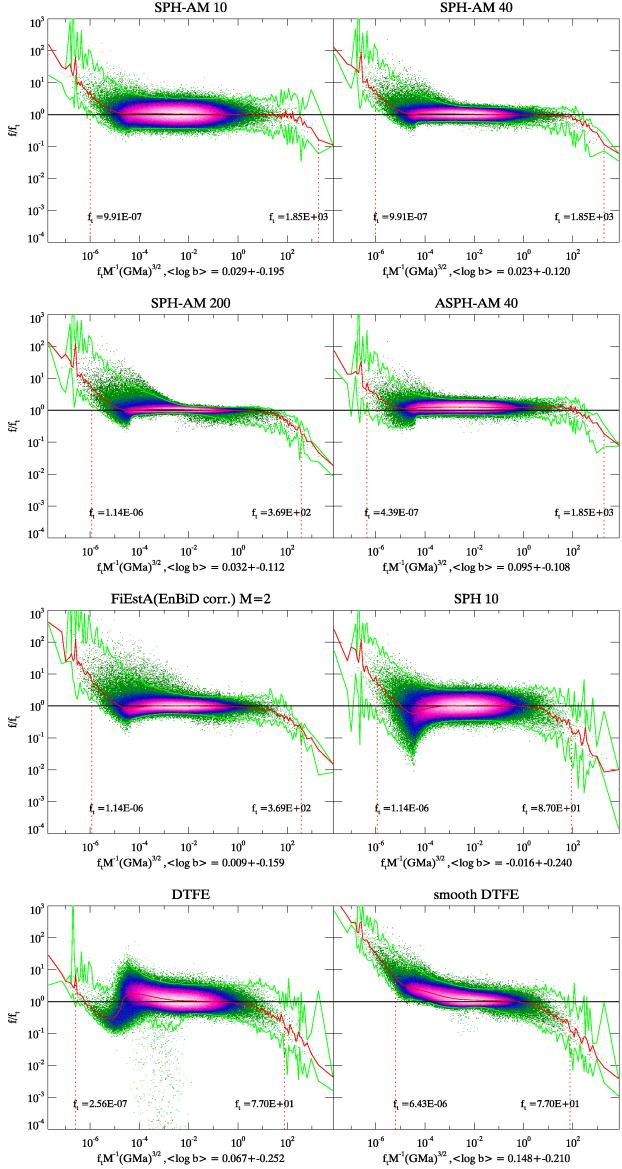}
\caption[]{The ratio $\ff=f/f_t$ between the measured phase-space
distribution function, $f$, and the analytical value, $f_t$,
as a function of $f_t$, derived for an isotropic Hernquist profile and different smoothing
methods as indicated on the top of each panel. 
{\sl From left to right and top to bottom}, (a) SPH with
local Adaptive Metric and 10 neighbours, (b) SPH
with local Adaptive Metric and 40 neighbours, (c) SPH with local Adaptive Metric
and 200 neighbours, (d) Anisotropic SPH with local
Adaptive Metric and 40 neighbours, (e) FiEstAS algorithm
with EnBiD improvement and $m_p=2.0$, (f) SPH with 10 neighbours, (g) DTFE method and (h)
DTFE with spherical sphere smoothing. 
These two dimensional histograms are calculated using 400x400 logarithmic bins. 
The central curve corresponds to the median value of $\ff$ calculated over
200 logarithmic bins along the $x$ axis, taking into account only
bins containing 2 particles or more. The two additional curves
on each side show $3$ sigma errors estimated from the dispersion
above and below the median curve.  The two dashed vertical lines
mark the range for which the median departs by more than a factor
5 from $f_t$. The mean log-ratio $\langle \log \ff \rangle$ and its
dispersion,  $\sigma=\sqrt{\langle (\log \ff - \langle \log \ff 
\rangle)^2 \rangle}$, are indicated on each panel and were computed using all the data.}
\label{hern1_fig}
\end{figure}

\begin{figure}
\includegraphics[width=8.5cm,height=4.25cm]{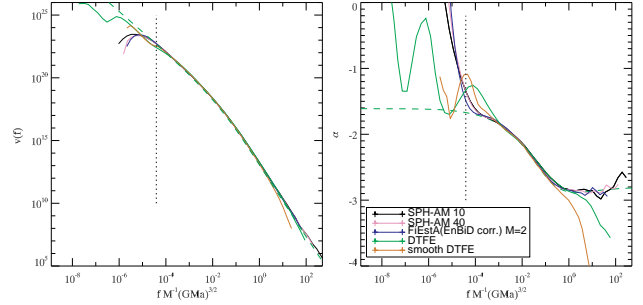}
\caption{Hernquist profile: measurement of $v(f)$ ({\sl left}) and
its logarithmic slope ({\sl right}) with different phase-space
estimators. The dashed line corresponds to the analytical
solution. The vertical dashed line marks the value $\log f_{\rm min}=-4.41$,
corresponding to the cut-off of the halo at $5$ virial radii.}
\label{hern3_fig}
\end{figure}

In order to test the various above described methods, we
first examine control ``phase mixed" samples  for which there are analytical
solutions\footnote{clearly, for such very symmetric  relaxed models with explicit first integrals, 
the best phase-space estimator would involve moving to angle-action variables and 
making use of the fact that the distribution function should not depend on the angles; 
since our purpose is to estimate phase-space density in more realistic settings this venue is not explored here.
Note nonetheless that the validation is carried here in this regime, which strictly speaking does leave open
discrepancies for a very unmixed system.
}.
 Hence, we follow A04 and S06 and
generate a Hernquist isotropic profile \cite{Hernquist1990}.
In that case, the projected density distribution is given by
\begin{equation}
\rho(r) = \frac{1}{2\pi3}\frac{M}{(r/a)(1+r/a)^3},
\end{equation}
where $M$ is the halo total mass and $a$ is a scale length. 
We follow exactly the prescriptions of A04 and S06 to create a random 
set of positions and random velocities obeying
the appropriate distribution,
relying on the fact that in this model, the phase-space density
distribution function, $f$, depends only on energy $E$,
\begin{equation}
E 
=\frac{v^2}{2}+\phi(r)=\frac{v^2}{2}-\frac{GM}{a}\frac{1}{1+r/a},
\end{equation}
where $r$ and $v$ correspond to position and velocity, respectively.
At equilibrium, the distribution function reads
\begin{equation}
f_t(E) = M
\frac{3\sin^{-1}(q)+q\sqrt{1-q^2}(1-2q^2)(8q^4-8q^2-3)}{4a^3\pi^3(2GM/a)^{3/2}(1-q^2)^{5/2}},
\label{eq:ftheoric}
\end{equation}
with
\begin{equation}
q=\sqrt{-\frac{E}{GM/a}}.
\end{equation}
In order to have a halo with realistic properties, we would like it to follow
equation~(\ref{EQ10}), i.e. $v_{\rm vir}=S_I r_{\rm vir}$ (in our case $r_{\rm
  vir}\simeq r_{200}$), with $S_I=1.0h\mbox{ km}~\mbox{s}^{-1}~\mbox{kpc}^{-1}$.
The circular velocity of the Hernquist profile reads
\begin{equation}
v_{\rm cir}(r) = \frac{\sqrt{GMr}}{r+a}.
\end{equation}
Combining this equation taken at the virial radius with equation~(\ref{EQ10}) 
gives the total halo mass
\begin{equation}
M=\frac{h^2 a^3 c (c+1)^2}{G}.
\end{equation}
In practice, the profile is also cut-off at a radius $r_{\rm cut}$, i.e. all the
particles verify  $r<r_{\rm cut}$. In what follows, a concentration parameter $c$ is defined
such that $r_{\rm vir}=c a$, where $r_{\rm vir}$ is the virial radius.

For our test sample, we take $5\cdot 10^5$
particles, $r_{\rm vir}=320$ kpc, $r_{\rm cut} = 5 r_{\rm vir}$, $h=0.7$ and $c=4$. 
For both the Hernquist profile and the Hernquist composite profile (next subsection),
we measure $f$ in units of $M^{-1}(GMa)^{3/2}$.
This choice of parameters was meant to compare directly our
measurements with S06. Note however that our value of $r_{\rm cut}$ is much
smaller than that of  S06, to make DTFE method tractable. Without such a
cut-off, we would indeed have too many neighbours for the particles
near the edge. This abrupt cut-off might a priori introduce
some contamination for $f \la f_{\rm min}$,
where $f_{\rm min}$ is the value of the phase-space distribution
function at $r=r_{\rm cut}$ and $v=0$ ($\log f_{\rm min}=-4.41$ in our
units). Yet, we have checked, by taking very large
values of $r_{\rm cut}$ that our implementation of the 
EnBiD estimator is quite consistent with that of  S06.

Figure \ref{hern1_fig} shows the ratio $\ff=f/f_t$ between the estimated phase-space 
density $f$ and the analytical result, $f_t$ given by equation~(\ref{eq:ftheoric}), as a function
of $f_t$, for various smoothing methods, as indicated on the top of each panel.

For SPH-AM with $10$ neighbours, we get a 
good approximation of $f_t$ over about $9$ decades delimited
by the two vertical dashed lines. These  lines  correspond to a five fold relative error 
on the determination of the phase-space density
compared to the exact solution. With $40$
neighbours, the spread drops down 
by almost a factor of $2$, but at the cost of a narrower
available dynamic range, because of  the bias
introduced by the softening  of the sharp transitions
between overdense and underdense regions and nearby local extrema
(see also the discussion in \S~\ref{sec:mysmoothing}). This effect is therefore even more
prominent for SPH-AM $200$.  The
FiEstAS algorithm with EnBiD improvement and $m_p=2.0$ gives a spread
comparable to SPH-AM with around $20$ particles, but the range of accurate phase-space
estimation is smaller as there is a noticeable bias
in the high density region. For Anisotropic SPH-AM, $64$ particles are
used to find the best fitting local ellipsoid while
softening is performed over $40$ particles.  
In this case, the plot looks almost the same as for SPH-AM with a
small overall systematic
overestimation of the true phase-space density, which
remains despite the kernel bias correction.

\begin{figure*}
\includegraphics[width=12cm]{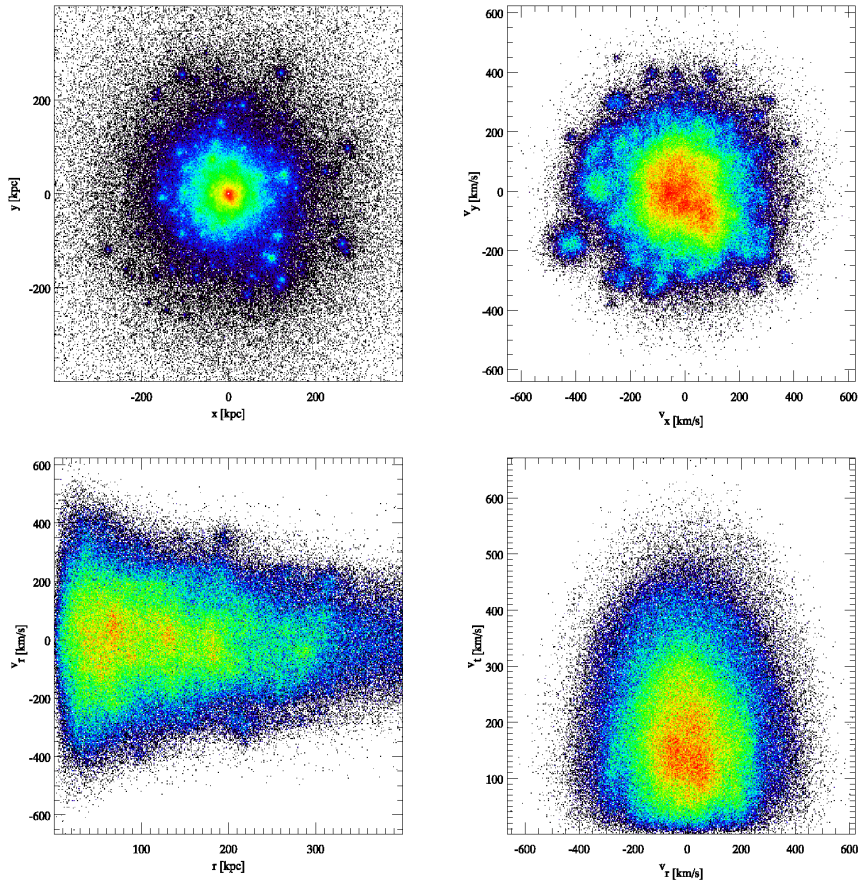}
\caption{the Hernquist profile with substructures in phase-space.
{\em Upper left panel:} $x$--$y$ position space; {\em upper right
panel:}  $v_x$--$v_y$ velocity space; {\em lower left panel:} phase-space
  diagram, radius $r$--radial velocity $v_r$; {\em lower right panel:}
radial velocity $v_r$--tangential velocity $v_t$. 
  Each plot is a two dimensional histogram with $400 \times 400$ bins. 
 Only the central part of the halo is shown here. }
\label{figure1a}
\end{figure*}

For DTFE and standard SPH methods, we have to set the global
position-velocity scaling factor $S_{\rm G}$, 
where $v=S_{\rm G} r$. We use the two methods described  in part (i)
of \S~\ref{sec:solsca}. The first one gives $S_{\rm G}=S_I=1.0h\mbox{ km}~\mbox{s}^{-1}~\mbox{kpc}^{-1}$ 
while the second one, relying on 
peak matching of the distance distribution, gives $S_{\rm G}=S_{\rm dist}=0.4h\mbox{ km}~\mbox{s}^{-1}~\mbox{kpc}^{-1}$. 
For both standard SPH and DTFE, we 
find that $S_{\rm dist}$ leads to a small but noticeable
improvement of a few percent for the phase-space density estimate compared to
$S_I$.
As shown on Figure~\ref{hern1_fig}, the SPH method with the $S_{\rm dist}$ scaling 
and 10 neighbours performs less well without the adaptive metric
correction, although it recovers correctly the middle range of $f$
values. Note that, contrary to the SPH-AM case, no edge effect 
correction is performed for the pure SPH case, which
explains the small depression seen at $f\simeq 10^{-4.5}$, due
to the cut-off at $r_{\rm cut}$.

Turning to the DTFE method, a simple estimation given by equation~(\ref{eq:DTFEsimple})
is closest to SPH with 10 neighbours, in terms
of scatter. However, within our (generous) allowed factor of 5 margin for
the phase-space density, we observe that the DTFE method probes the low-$f$
range nearly one order of magnitude further, while
it seems to do worse in the high density regime. Note 
the bump at $f \simeq 10^{-4}$ seen in the lower left
panel of Figure~\ref{hern1_fig} in addition to the neighbouring
depression already noticed for SPH, here at $f \simeq 10^{-5}$. 
It is a consequence of the brutal cut-off at $r_{\rm cut}$, combined with the strong
effects of anisotropy in phase-space near the edges: indeed, for
$f \la 10^{-5}$, the number of neighbours defined
by the Delaunay cells starts to increase dramatically
(see Figure~\ref{con} in \S~\ref{sec:smoothneig}).

We checked alternate smoother DTFE interpolators discussed in \S~\ref{sec:DTFE}, and 
found that the best one is the ``spherical'' smoothing implementation given by equation~(\ref{eq:DTFEsph}).
This solution, shown on Figure~\ref{hern1_fig} presents less scatter than the
simple DTFE and a slightly better behaviour with respect
to edge effects, at the cost of a significant reduction of the available
dynamic range, which covers only about 7 decades.

Another way to test our estimators, following Arad04,  involves measuring the
probability distribution function of $f$, which is,
 within a normalisation factor, the differential volume
\begin{equation}
v(f)=\frac{{\rm d}V}{{\rm d}f},
\end{equation}
where $V(f_0)$ is the volume in phase-space occupied by the excursion
$f > f_0$:
\begin{equation}
V(f_0) = \int_{f_0}^\infty v(f') {\rm d}f' = \int_{f(x,v)>f_0}{\rm d}^3x{\rm d}^3v.
\end{equation}
For an isotropic Hernquist profile, the function $v(f)$ can again be
computed analytically (see \S~3.1 of S06 for details). 

The measurement of $v(f)$ is straightforward when one considers
the simplest implementation of DTFE as $V(f_0)$ is given exactly
in that case by
\begin{equation}
V(f_0) =\sum_{f_v>f_0}\frac{m_p}{f_v}.
\label{eq:voff}
\end{equation}
For other methods, equation (\ref{eq:voff}) is only approximate.
The logarithmic derivative of function $V$ is then obtained by simple finite
difference in $\log f$ space using $100$ bins, using 3 points
interpolation. 

Following Arad04,  let us also estimate the 
logarithmic slope, $\alpha$, of the function $v(f)$,
\begin{equation}
\alpha(f) = \frac{{\rm d}\log[v(f)]}{{\rm d} f},
\end{equation}
since it represents a more discriminant measure of the phase-space density 
than $v(f)$ itself.

This is illustrated by Figure~\ref{hern3_fig}, which compares to
the analytical solution the measured $v(f)$ and its logarithmic slope.  
Note that, because of our cut-off at $5.0 r_{\rm
  vir}$, $v(f)$ (and therefore its logarithmic derivative) are not expected
to fit the analytical prediction for $\log f \la \log f_{\rm
  min}=-4.41$, since a fraction of the sample volume $V(f)$, is missing
in that regime. All the methods reproduce quite well $v(f)$ and 
$\alpha(f)$ above that value and in the mid-density regime. In the high
density regime, the best results seem to be obtained by SPH-AM with 10
particles, but the measurements are too noisy to be definite:  
one can see that SPH-AM with 40 particles and FiEstAS
with EnBiD correction do as well at least for $f \la 1$. On
the other hand, the DTFE method behaves quite
poorly in the high density regime, while its smoother counterpart is
even worse, which confirms the results
of Figure~\ref{hern1_fig}. 
However, we shall see that this is a consequence of
a suboptimal choice of the scaling between position and velocities,
as discussed in next section. Actually, with the proper choice of
$S_{\rm G}$, DTFE should give the best results in high density
regions, as, by construction, it provides a full tessellation of space 
with optimal calculation of neighbours: the combination of these last 
two
properties is critical to measure accurately 
an Eulerian quantity such as $v(f)$. 

\subsection{Hernquist composite profile}
\label{sec:hernquist_composite}

Our single isotropic Hernquist profile allowed us to separate well
the different density regimes. However it is not realistic, since real
dark matter haloes exhibit non trivial substructures. 
We therefore now create a synthetic halo with a main component and
smaller subhaloes. We still use the Hernquist profile as a guideline
to be able to perform analytical predictions. 

For the main halo we use the same realisation as before with around 
$2.5 \times 10^5$ particles. Then  we add $500$ smaller haloes 
which correspond to a scaled-down version of the main halo. 
Their mass follow the following probability distribution function, 
$p(M){\rm d}M\propto M^{-1.8}{\rm d} M$, where $M$ varies between $M_{\rm min} =
0.00025M_{\rm main}$ and $M_{\rm max}=0.06M_{\rm main}$. The largest subhalo has around 
$14,000$ particles and the smallest one, around $60$. 
In total, the system involves, as before, $5.0 \times 10^5$
particles, and around $50\%$ of them belong to substructures. 
This fraction is larger than what is  found  in cosmological simulations, but we prefer 
this ratio given its higher level of anisotropy. Each subhalo
phase-space coordinates centre is set randomly following
the same Hernquist distribution as for the main halo.

Figure \ref{figure1a} presents the halo in various
projections. As illustrated by the top panels, we 
can see clearly that the structures are more
concentrated in position space than in velocity space, a feature  also observed in $N$-body
haloes (see for instance Figure~\ref{figure2bb}).
The bottom panels correspond to phase-space diagrams,
in radius/radial velocity subspace (lower-left panel) and
in radial velocity/tangential velocity subspace
(lower-right panel). To draw 
them, we compute for each particle the distance $r$ from the centre
of the main halo and the relative radial velocity as follows:
\begin{equation}
v_r = \frac{1}{r}\sum_i(x_i-x_{c,i})(v_i-v_{c,i}), 
\end{equation}
where $i=1,\cdots,3$ corresponds to the coordinate,
while $x_c$ and $v_c$ are the position and velocity  of the 
centre of the main halo, respectively. 
The tangential velocity is then given by $v_t=\sqrt{(v-v_c)^2-v_r^2}$. 
Note the elongated vertical features in lower-left panel, which
illustrate again the smaller spread of substructures in position space
than in velocity space. 

\begin{figure*}
\includegraphics[width=15cm,height=5cm]{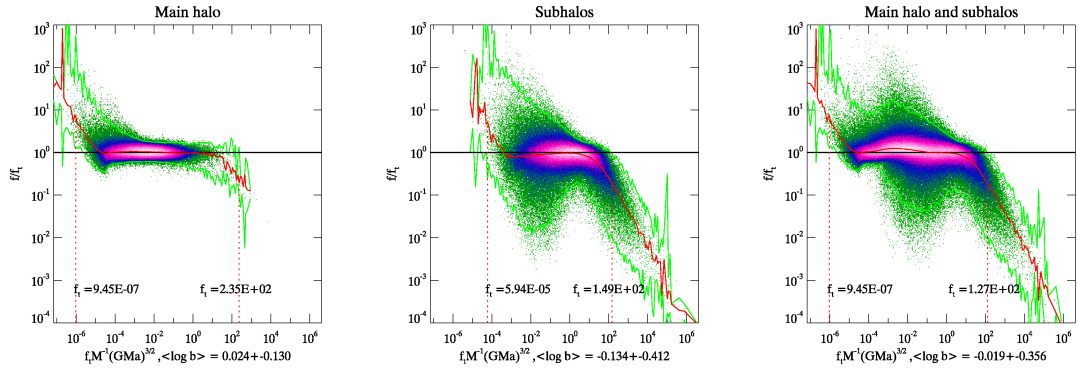}
\caption[]{Comparison with the analytical prediction of 
numerically estimated phase-space density from various
  contribution of our composite Hernquist halo. The measurement is
  performed with SPH-AM method using 40 neighbours. The quantity
  displayed is $f/f_t$ as a function of $f_t$, where $f$ and $f_t$ are
  the measured and the analytical phase-space distribution functions,
  respectively.  {\em Left panel:}
  main component only. {\em Middle panel:} subhaloes, only. In that
  case the ratio $f/f_t$ is computed for each subhalo, individually.
  {\em Right panel:} the full profile, with haloes and subhaloes,
  while $f_t$ is given by the sum of each analytical profile
  contributing locally.}
\label{figure1ens}
\end{figure*}
\begin{figure}
\includegraphics[width=8.5cm,height=17cm]{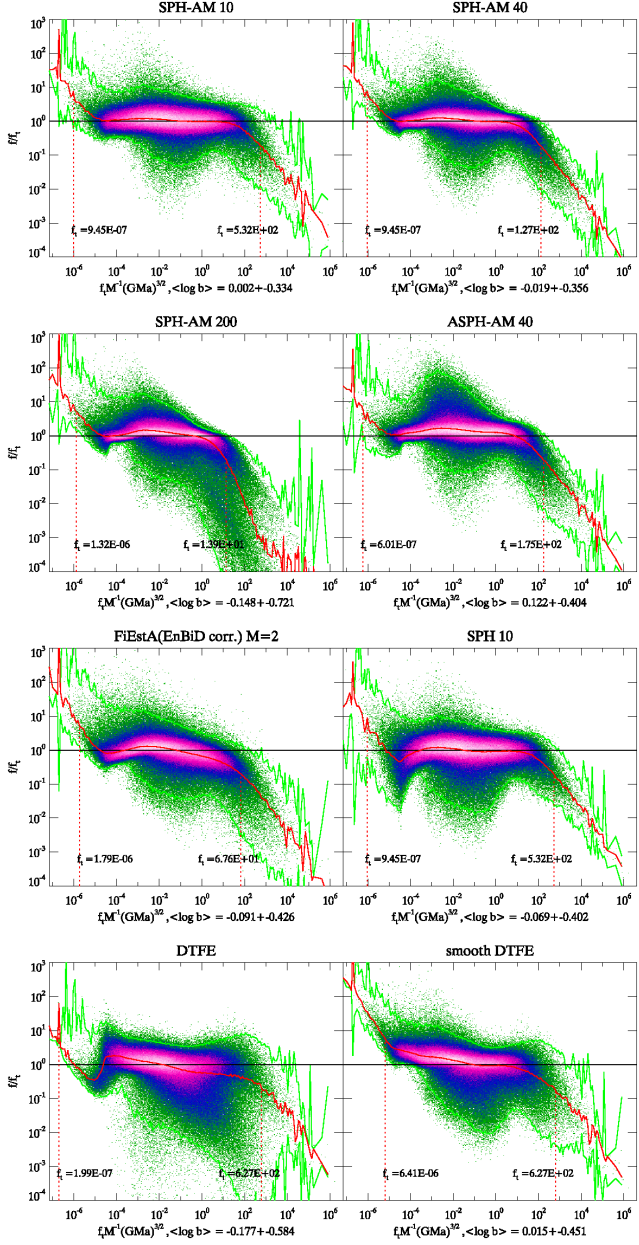}
\caption[]{Same as in Figure~\ref{hern1_fig}, but the ratio between
measured and analytic phase-space distribution function is now shown
for our composite Hernquist profile for various
smoothing methods as indicated on the top of each panel. }
\label{figure2}
\end{figure}

\begin{figure*}
\includegraphics[width=15cm,height=5cm]{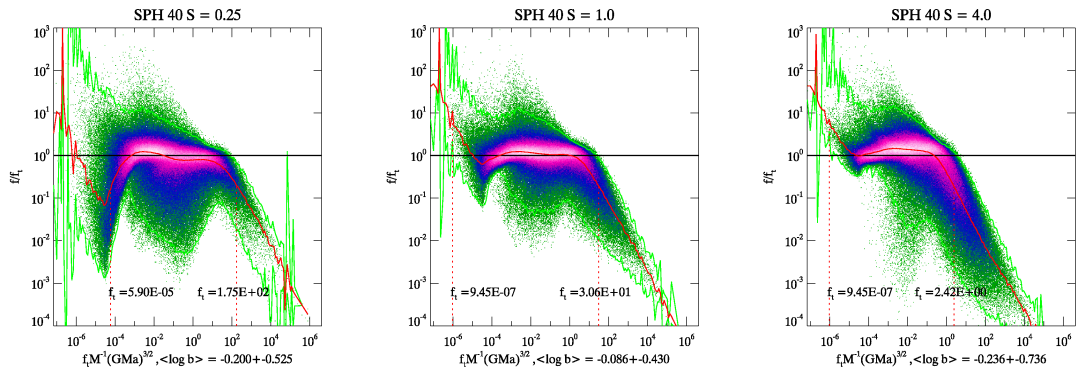}
\caption[]{Effect of the choice  of the scaling parameter $S_{\rm G}$
between positions and velocities. The  ratio $f/f_t$ is shown
as a function of $f_t$, following  Figure.~\ref{figure2}, but
for SPH method with 40 neighbours. {\sl From left to right},
$S_{\rm G}=0.25h\mbox{ km}~\mbox{s}^{-1}~\mbox{kpc}^{-1}$,  
$S_{\rm G}=1h\mbox{ km}~\mbox{s}^{-1}~\mbox{kpc}^{-1}$, $S_{\rm G}=4.0h \mbox{ km}~\mbox{s}^{-1}~\mbox{kpc}^{-1}$.}
\label{figure3}
\end{figure*}

\begin{figure}
\includegraphics[width=8.5cm,height=8.5cm]{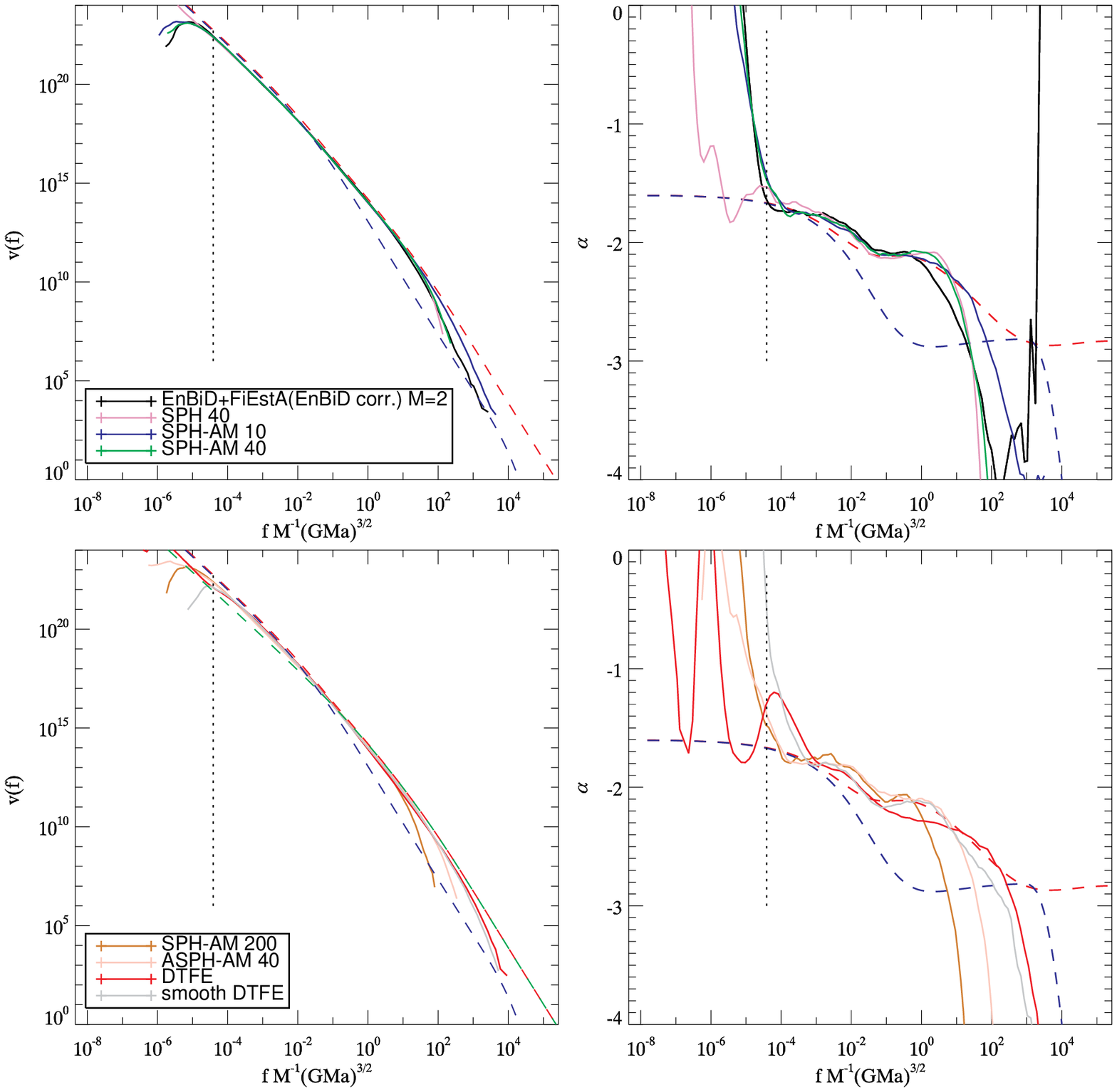}
\caption{Measurement of $v(f)$ and its logarithmic derivative,
  following Figure~\ref{hern3_fig}, but for the composite
  Hernquist profile. The blue and red dashed curves correspond to the
  analytical prediction for the main component and
  the full halo, respectively.
  Note that, as discussed in \S~\ref{sec:hernquist},
  there is a minimum value of $f$ for which we can measure accurately
  $v(f)$, regardless of the method used, due to the cut-off imposed at radius
  $r_{\rm cut}$ (dashed vertical line). This is now further complicated by the fact that here,
  the cut-off is also imposed on the subhaloes, which explains
  why the measurements tend to slightly overestimate the
  red dashed curve for $10^{-4} \la f \la 0.1$.}
\label{figure5}
\end{figure}

Figure \ref{figure1ens} shows the phase-space density estimated
by SPH-AM with 40 particles, for the main component, the subhaloes
and the full halo. While the theoretical density is a simple
sum of all the components contributing locally, it is not exactly the
case for the estimated density. By comparing all the plots, one
can see that the low-$f$ regime and the high-$f$ regime are dominated
by the main component and the subhaloes, respectively. For
each component, $f$ presents a large spread  
in the low density region, $10^{-4} \la f \la
1$, because of the high level
of shot noise due to the small number of particles in the edges
of substructures, and 
is underestimated in the high density one.  The range of accurate values
of $f$ increases with the number of particles in each subhalo. When
summing up the subhaloes, as shown in middle panel,  
this adds up to a significant spread of the scatter plot, obviously much
larger than for the main component. In fact, such a spread
dominates for the total halo (right panel) and is larger than for 
a single Hernquist profile with same number of particles (see
Figure~\ref{hern1_fig}). 
Because the high phase-space
density regime is dominated by subhaloes, the range of recovered values
of $f$ is tremendously reduced in that region, and 
we lose about  one order of magnitude for the available
high density range compared to the single Hernquist profile. 
This issue has to be kept in mind when performing the measurements in
real $N$-body haloes.

Figure \ref{figure2}, following Figure~\ref{hern1_fig}, now compares the various
estimators of the phase-space density, and confirms most of our
previous findings: the best estimator is 
SPH-AM  with 10 neighbours. It does better than SPH-AM
40 in the high-density regions, because of  the lower-level of smoothing,
at the cost of a larger spread. The effect is even stronger when
performing the comparison with SPH-AM 200, as expected. Again,
ASPH-AM with 40 neighbours seems to bring some global estimation bias.
EnBiD-FiEstAS with $m_p=2$ does not perform any better than SPH-AM since
it seems to underestimate  $f$ earlier in the high density regime.

Regarding the global scaling, the findings of Figure~\ref{hern1_fig} are
confirmed: we obtain the best results by matching the
nearest neighbour distance distributions 
and we measure again (but this coincidence
is not generic) $S_{\rm dist}=0.4h\mbox{ km}~\mbox{s}^{-1}~\mbox{kpc}^{-1}$). 
The suboptimal nature
of the global scaling induces  an  increase of the amplitude of the scatter
below the median, for instance when one compares SPH-AM 10 with
SPH 10 on Figure~\ref{figure2}. Turning to DTFE in its basic
implementation, which stills perform best for
low values of $f$ (except for the irregularity already observed
in Figure~\ref{hern1_fig}), this spread becomes dramatic and
strongly asymmetric but can be reduced
by using the smoother and more isotropic ``spherical'' interpolator
$f'$.
Note that, given our factor of five tolerance between measured
and exact phase-space density distribution function,
DTFE and its smoother version do better than in Figure~\ref{hern1_fig}. 
In fact they now seem to perform slightly better than SPH-AM 10 in the 
very high-density regime. Indeed,  the fraction of
overdense particles intervening in the calculation
of $S_{\rm dist}$ is much larger: the calculation of $S_{\rm dist}$,
corresponding to a compromise between all the particles, is now more adapted to
the overdense part of the phase-space distribution function.
For the single Hernquist profile, the fraction of particles belonging
to the high $f$ part was indeed much smaller. 
Hence, provided that the proper 
global scaling between velocities and positions has been set up,
DTFE chooses by construction the proper adaptive smoothing range (or
the right number of neighbours). However note that the overall shape of the median
curve of Figure~\ref{figure2} is not as flat as for SPH-AM, and this
is a consequence of the fact that the global scaling is only a compromise
that is not locally optimal. In fact, in addition to the small-$f$
irregularity already observed in Figure~\ref{hern1_fig}, the high-$f$ plateau in
lower-left panel of Figure~\ref{figure2} is somewhat
below the thick horizontal line. This  follows from the presence of
substructures, as discussed above, which does not only induces a strong asymmetry
of the spread around the median value: it also biases it
to lower values. This is
because DTFE uses many neighbours in that regime to perform the
interpolation, (about 200 as will be discussed in next section,
see Figure~\ref{con}), which makes it very
sensitive to  the local anisotropies in the phase-space distribution
function. The bias on the high-$f$ plateau 
is at least partly corrected for by the ``spherical'' version of
DTFE, which is indeed expected to less sensitive to such anisotropies, as
illustrated by lower-right panel of Figure~\ref{figure2}.

To illustrate in more detail the influence of the choice 
of the global scaling parameter, $S_{\rm G}$, between velocity space and
position space, Figure  \ref{figure3} shows, for our composite
profile, how the quality of the measurement
of $f$ changes with $S_{\rm G}$ for the SPH method with 40
neighbours (similar trends would be seen for the DTFE method, while
adaptive metric methods, e.g. SPH-AM with 40 particles,
are by definition totally insensitive to the choice of $S_{\rm G}$). 
The domain of valid estimates  for $f$
considerably depends on the choice of $S_{\rm G}$, as a change
by a factor 4 in $S_{\rm G}$ induces a loss by an order of
magnitude in the high density range. Note in particular that the shape of the scatter,
below the median curve, changes with the actual value of $S_{\rm G}$. For
instance, this scatter is reduced in the intermediate range of values for
$f$ in the middle panel of Figure~\ref{figure3}. This is again a consequence
of the fact that the optimal value of $S_{\rm G}$ depends on location
in phase-space, in particular on the local distribution of
velocities versus positions in the core of substructures, i.e.
in the neighbourhood of local density peaks in phase-space.

Following   \S~\ref{sec:hernquist}, let us finally turn
to the measurement of $v(f)$ and its logarithmic
derivative, $\alpha(f)$, as illustrated by Figure~\ref{figure5}.
The analytic calculation of $v(f)$ is similar to
\S~\ref{sec:hernquist} except that we now consider each component
separately and then combine them straightforwardly.
The measurement $v(f)$ itself is performed  exactly as
explained in \S~\ref{sec:hernquist}. The results of
Figure~\ref{figure2} are partly confirmed by Figure~\ref{figure5}: the best adaptive
metric method is again SPH-AM with 10 neighbours. However the
best measurements are by far now given by the basic 
DTFE method (without additional
smoothing). 
Recall that this is due, in part, by the fact that
the calculation of $v(f)$ is better behaved for the DTFE method
than for other methods: indeed the concept of Eulerian volume
is well defined for DTFE, while it remains only approximate with the
SPH methods. These methods are optimal when one sits on particles, but
get more and more inaccurate when one goes away from the particles.
In that sense, Figure~\ref{figure2}, which uses a pure Lagrangian point
of view, greatly  favours the SPH methods, while the measurement of
$v(f)$, which is intrinsically an Eulerian quantity, favours more
DTFE.

Globally, the measurements in this section  suggest that the
DTFE method performs rather well, provided that the correct position/velocity
scaling is set up.  However, the very calculation of
the correct value of the scaling factor, $S_{\rm G}$, is not straightforward: 
in \S~\ref{sec:hernquist}, the DTFE method was performing
poorly.  Even if it is well estimated,
this global scaling provides only a compromise, that is not locally
optimal.

Finally, let us mention some additional issues about the DTFE method.
While exploring various
values of $S_{\rm G}$, we found that with larger $S_{\rm G}$,
this method starts to generate very large number
of Delaunay cells which becomes rapidly impossible to handle
computationally. The same happens when we increase the cut-off
radius $r_{\rm cut}$: in that case, particles
near the border of the catalogue present tremendously large
number of Delaunay cells, as they are connected to almost all the
other particles. Indeed, it is expected that a high level of local anisotropy
increases the number of DTFE neighbours.\footnote{Note that
a way to compute the optimal value of $S_{\rm G}$ could
involve minimising the total number of Delaunay cells.} 
For all these reasons, we favour the
SPH-AM method relative to the  DTFE method, even if they seem to perform
less well for the measurement of Eulerian quantities such as $v(f)$. 
Still,  if one put aside the problem of
position/velocity scaling, the DTFE method provides a local estimate of the optimal number
of neighbours, which can help to find the best number of neighbours
for the SPH-AM method, as we shall discuss in \S~\ref{sec:smoothneig}.
Finally, while potentially better
than the SPH-AM method, the  ASPH-AM methods tend to yield a slight overestimation
bias for the mid-range of values of $f$, and we did not find
a straightforward way to correct for it.

\subsection{Haloes from N-body simulations}
\label{sec:Nbodyhalo}

\begin{figure*}
\includegraphics[width=12cm]{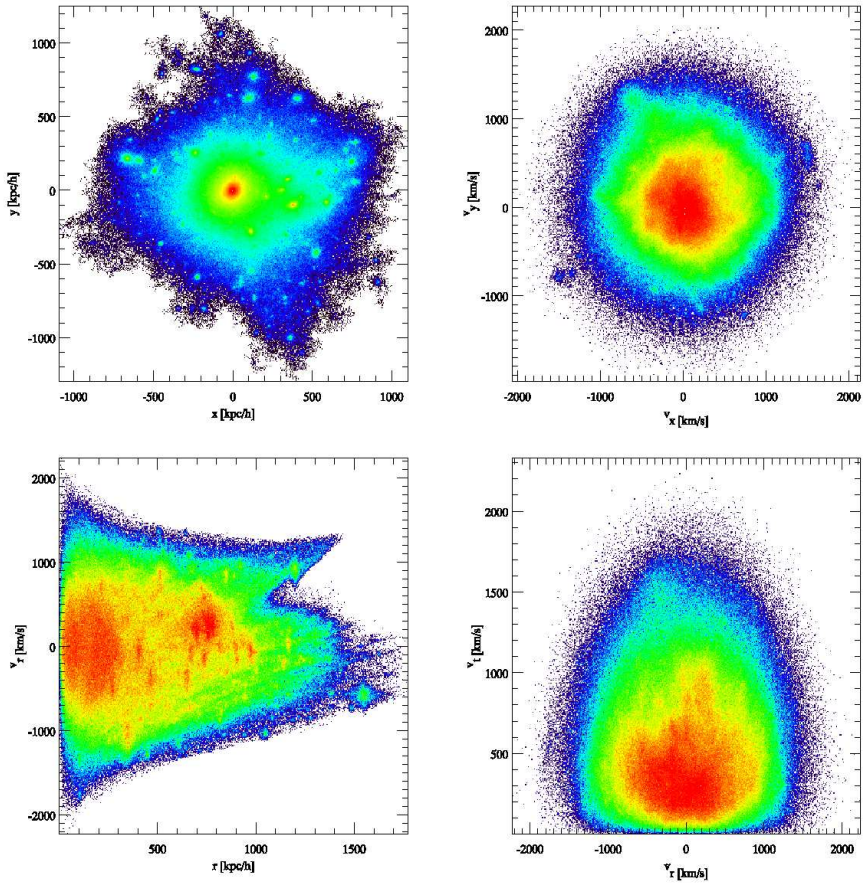}
\caption{Appearance of our CDM $N$-body halo with 1.8 million
particles. This figure can for instance be compared to Figure~\ref{figure1a}.
{\em Upper left panel:} $x$--$y$ position space; {\em upper right
panel:}  $v_x$--$v_y$ velocity space; {\em lower left panel:} phase-space
  diagram, radius $r$--radial velocity $v_r$; {\em lower right panel:}
radial velocity $v_r$--tangential velocity $v_t$.}
\label{figure2bb}
\end{figure*}

We now consider the
realistic case of a halo extracted from a Cold Dark Matter (CDM)
$N$-body simulation. To do that, we performed a standard CDM simulation
with GAGDET2 \cite{Springel2005} involving $512^3$ particles in a periodic cubic box
of size $50 \ h^{-1}$ Mpc. 
The choice of the cosmological
parameters is matter density $\Omega_M = 0.3$ and cosmological
constant $\Omega_\Lambda = 0.7$. The linear variance of the
density fluctuations in a sphere of radius $8 \ h^{-1}$ Mpc is
$\sigma_8=0.92$ and the Hubble
constant fixes $h=0.7$. This is slightly
different from recent constrains e.g. provided by WMAP \cite{Spergel2003}
 but should be close enough for our purpose.
For reference, these cosmological parameters fix the mass of a
particle to be $7.7\times 10^7  M_{\sun}$. Haloes were extracted at present
time from this simulation using standard FOF algorithm with linking parameter
$b=0.2$. To make the calculations tractable for DTFE, we selected
 the third most massive halo, which contains about
$1.83\times 10^6$ particles. Only the linked particles are considered.
In the subsequent analyses, calculations
are performed in comoving phase-space coordinates instead of physical ones. 
However, when it comes to phase-space density calculation, 
the main change when passing from one system of coordinates to the other comes
from the effect of the Hubble flow, which has rather insignificant impact on 
the final results.

Figure \ref{figure2bb} displays various
projections of the halo, following Figure~\ref{figure1a}. Here,  only
one additional complication arises in order  to calculate
correctly phase-space diagrams: the centre of the halo has to be
defined accurately in phase-space. We find this centre through
an iterative process, applied to each subspace separately.
The first step involves considering the centre as the mean
position (velocity) of all the particles. Then, the distance of each particle
from this centre is computed and half of the particles are
removed by choosing the most distant ones. A new centre is computed
from the remaining particles. The process is repeated again as long
as there are more than 100 particles left. 

As noted earlier,  the velocity subspace (upper panels of
Fig.~\ref{figure2bb}) is relatively featureless. 
Figure \ref{figure2bb} is
in fact very similar to Figure~\ref{figure1a}, except 
for the lower-left  panel which displays
more complex structures. In particular,
in addition to the vertical ``fingers'', one can notice some
elongated structures  that correspond to non trivial filamentation
of phase-space built up by the dynamics, e.g. tidal tails (see for instance
Peirani \& Pacheco 2007).

\begin{figure}
\includegraphics[width=8.5cm,height=17cm]{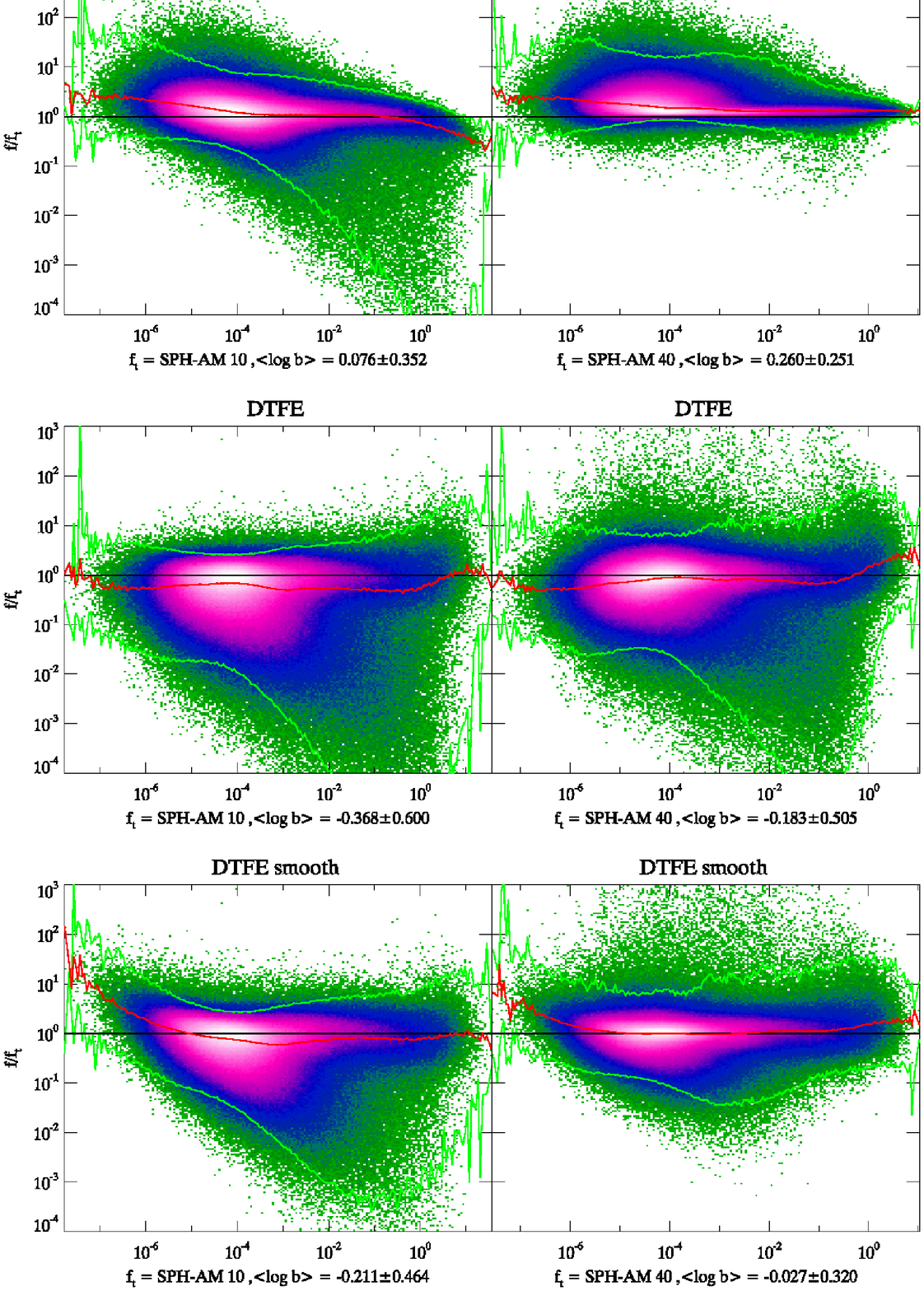}
\caption{Phase-space density estimation for our
CDM $N$-body halo with $1.8$ 
million particles. The left and right columns
correspond to  the  SPH-AM method with 10 and 40 neighbours, respectively,
for the theoretical phase-space density, $f_t$.  
From top to bottom, the smoothing method considered is
(a) the SPH-AM  method  with 40 neighbours on the left panel and SPH-AM  with 10 neighbours
on the right panel, (b) the  ASPH-AM with 40 neighbours on the left and
ASPH-AM with 10 neighbours on the right panel, (d) the basic DTFE method
and (e) DTFE method with spherical smoothing. 
}
\label{figure10}
\end{figure}

In this more realistic framework,
we cannot rely on  an analytic expression of the reference $f_t$ to
perform plots similar to Figures~\ref{hern1_fig} and \ref{figure2}.
However since the SPH-AM methods have our preference, we shall now use
them as references. This is illustrated by Figure~\ref{figure10},
which shows the ratio  $f/f_t$ as a function of $f_t$ for various
smoothing methods; $f_t$
is given  this time  by the  SPH-AM methods with 10 and 40 neighbours, for respectively  
the left and right columns. Note that $f$ and $f_t$ are measured
for each particle individually to perform these
scatter plots.
To fix the global scaling position/velocity parameter for the DTFE
implementations, we use a   coincidence 
scaling  of the peak distance distribution given by $S_{\rm G}=S_{\rm dist}=0.38h\mbox{ km}~\mbox{s}^{-1}~\mbox{Mpc}^{-1}$.

Figure \ref{figure10} confirms qualitatively the results found for
the single and the composite Hernquist profiles. In particular, the  SPH-AM methods
with 40 neighbours underestimates high phase-space densities compared 
to  the SPH-AM 10 method. In the upper left panel of Figure~\ref{figure10}, 
there is also a tail below the median curve, 
which arises because the measured $f$ is 
significantly biased toward lower values nearby local phase-space density peaks 
corresponding to each substructure, 
as explained in previous section; this local bias is more prominent for the 
SPH-AM 40 than SPH-AM 10 method, and consequently the median curve of the 
upper left panel is slightly below unity, except for low-$f$. 
This bias can be reduced with adaptive smoothing,
as shown in the second row of Figure~\ref{figure10} with the 
ASPH-AM 40 method. But recall that this is  achieved at a price of a slight
uncontrollable positive bias, here in underdense
phase-space regions. The  DTFE method seems to behave well overall, with
a positive bias in underdense phase-space regions when implementing
its smoother 
version. However, our appreciation is again skewed
by the somewhat loose factor of five tolerance.
In fact, DTFE in its simpler implementation 
seems to globally underestimate the phase-space
density distribution function, except in the very-low and
in the very-high density regimes. Again, this is due to the
contribution  from substructures, which
is now more significant given the larger effective number of
neighbours used by DTFE and its high sensitivity to the choice
of the local scaling, $S_{\rm L}$. 
The effect of the tail below the median value is therefore
now more significant than for the upper left panel, and it is 
reduced, as well as the bias of the median curve in intermediate values 
of $f$, by the spherical
implementation (lowest panels of Figure~\ref{figure10}).
Hence Figure \ref{figure10} globally
confirms the findings of Figure~\ref{figure2}.
Note that we do not observe any irregularity in the low-$f$ regime as in 
Figure~\ref{figure2}, because the cut-off of the halo is 
performed in a much smoother way.

\section{Additional insights}
\label{sec:additiona}

Although the Delaunay tessellation cannot easily address the problem
of the  position/velocity scaling, because of
it self adaptive nature, it still provides some
insight about the local neighbourhood, in particular about
the optimal number of neighbours that should be used in SPH methods.
In \S~\ref{sec:smoothneig}, we analyse in details the neighbour distribution provided by the Delaunay
tessellation. This will help us to better  understand the results found in the previous
sections and to further justify our preference for the SPH-AM estimator with
a number of neighbours ranging from 10 to 40.

The analyses of \S~\ref{sec:numexp} show that
the entropy method implemented in EnBiD provides
a very good approximation of the local scaling  to apply between
positions and velocities. In \S~\ref{sec:posvelcor2}, 
we investigate how this scaling depends
on the environment, and in particular on the value of $f$. This will allow
us to  better understand to choice of the global scaling.

\subsection{Smoothing range and local neighbourhood}
\label{sec:smoothneig}

\begin{figure}
\includegraphics[width=8.5cm,height=8.5cm]{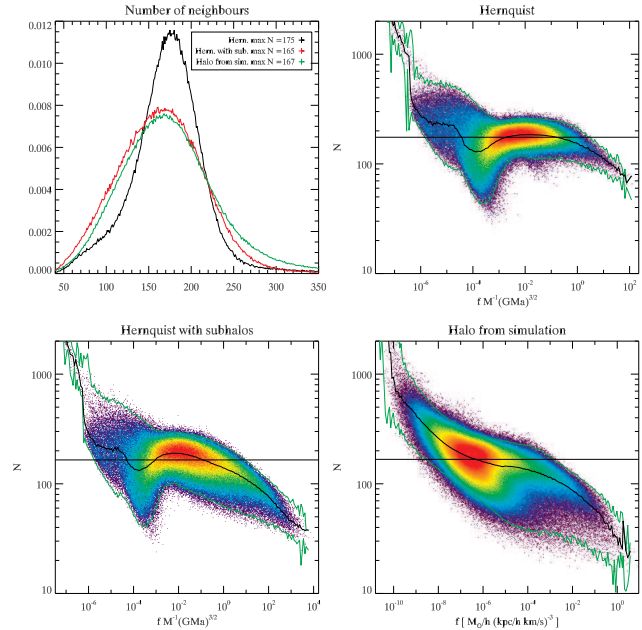}
\caption{Number of neighbours $N$ found by the Delaunay tessellation for
the three profiles considered in this paper. 
{\it Top left panel:} the probability distribution function, $P(N)$,
for a  particle of having $N$ neighbours. {\it Top right panel:} $N$ is shown as a
function of the theoretical phase-space density, $f_t$, for our single Hernquist
profile. The horizontal line corresponds to the mean value of $N$,
while the smooth black curve gives the median as a function of $f_t$.
{\it Bottom left panel:} $N$ is shown as a function of the 
theoretical phase-space density, for our composite Hernquist profile.
{\it Bottom right panel:} $N$ is shown as a function of the 
phase-space density measured in our dark matter simulated halo.}
\label{con}
\end{figure}

In Figure~\ref{con}, we study the distribution of the number of neighbours
built by the Delaunay structure as a function of phase-space density,
for the single Hernquist profile of \S~\ref{sec:hernquist} (upper
right panel), the composite Hernquist profile of
\S~\ref{sec:hernquist_composite} (lower left panel)
and the $N$-body halo of \S~\ref{sec:Nbodyhalo} (lower right panel).
The upper left panel shows, for each instance, the overall distribution function of the
number of neighbours. 

As expected, the average number of neighbours is approximately the same in the
three cases: $\langle N \rangle=175$, 165 and 167, for the Hernquist profile,
the composite Hernquist profile and the $N$-body halo, respectively. 
The presence of substructures widens the overall distribution of values of
$N$, as shown by the green and the red curves in upper left panel
of Figure~\ref{con}, as compared to the black one. 
In the three cases considered, the typical number of neighbours
decreases with increasing phase-space density, following three regimes: 
\begin{enumerate}
\item {\em particles near the edges:} at the edges of
the catalogue, where $f$ is very small, $N$ is very large, of the order
of 1000 for the Hernquist cases up to nearly 10\,000 for the $N$-body halo.
It then decreases rapidly, while particles are getting
away from the edges to reach the next regime, (ii). Note that 
$N$ being large is not a consequence of $f$ being small, but follows from
the fact that the phase-space distribution function 
presents an overall positive curvature and is highly anisotropic
because  of the edges.
\item {\em The plateau at intermediate
values of $f$, far from the edges and from the main
distribution of local maxima:} in this quiescent regime, we have
 $N \sim \langle N \rangle$, where $\langle N
\rangle$ is the typical number of expected Delaunay neighbours 
just as quoted above. Note that there is
a slight difference between the Hernquist profiles and the $N$-body
halo, in particular a lower bump at $f \simeq 10^{-4.5}$ in upper
right and lower left panels, that corresponds to the
transition between regime (i) and regime (ii), and
which does not appear for the $N$-body
halo. This is probably due to the brutal cut-off imposed at
radius $r_{\rm cut}$ as mentioned in \S~\ref{sec:hernquist},
which can affect the neighbour distribution in a non trivial manner  
up to $f\simeq 10^{-4.41}$ for the single Hernquist profile
and the main component of the composite Hernquist profile. 
\item {\em The high density regime, dominated
by the regions nearby local maxima:}
the number of neighbours decreases again rapidly because
$f$ now  presents an increasingly overall negative curvature when one
reaches the densest regions, which makes $N$ smaller; we
measure it to be as small as
30 in the presence of substructures (which dominate
large values of $f$, as noticed in \S~\ref{sec:hernquist_composite}), 
while it remains close to 100 in the single Hernquist profile.
\end{enumerate}
Intuitively these numbers suggest that, when turning to SPH
methods, the number of neighbours used to perform the interpolation
should depend on the environment. In particular, in the ``quiescent''
regime, i.e. far from the edges
and from the peaks, we should take around 200 neighbours to perform
the measurements. However, such a large number of neighbours is not
optimal near the peaks: the DTFE algorithm suggests a value of $N$ of the
order of a few tens for sampling best the core of substructures, which
is fully confirmed by the analyses of our previous sections.
Furthermore we noticed that these values of $N=10$ and
$N=40$ are still  appropriate in the quiescent regime: only the 
signal to noise ratio --the spread
due to local Poisson noise around the local average value-- is
changed. Taking a value
of $N$ as large as 200 provides too much smoothing and biases the
results in overdense regions. Moreover it also induces non
trivial diffusing mixing effects. The larger the number of
neighbours, the more sensitive the determination of $f$
to local anisotropies.

Such local anisotropies (and local
curvature properties) can be captured better --at least partly-- 
by anisotropic SPH smoothing (see the discussion in \S~\ref{sec:mysmoothing}
and Figure~\ref{sph_fig}), but at a risk of some potential
slight positive biases, as argued previously. 
Since SPH methods in their usual implementation are not self adaptive in terms of their number
of neighbours,\footnote{It should be however rather easy to implement
  SPH smoothing with a number of neighbours varying with the
  environment.} 
we think that the best choice for $N$ is a value ranging
from 10 to 40, because such a choice offers a good compromise for all the 
dynamic range. Note that this also confirms the findings of
S06. There is still the problem of what happens near the
edges, but these can sometimes be extended sufficiently far away to avoid
contamination of the measurements in the region of interest. What
really influences the results are abrupt changes of local curvature: 
while the DTFE method captures them optimally, 
SPH method does it only approximately, and
of course,  does it best when the number of neighbours remains small,
at the expense of a slightly worse local signal to noise ratio. 

Note, however, that this discussion is again biased by the fact that
SPH uses a Lagrangian point of view, i.e. it sits on
the centre of each particle to measure locally the phase-space
distribution function. An Eulerian point of view, that would require
to measure $f$ in an arbitrary point of space (see, e.g.,
Colombi, Chodorowski \& Teyssier 2007), would probably impose
a slightly larger number of neighbours for SPH methods 
to give sensible results.

Note finally 
that the findings of Figure~\ref{con} are of course sensitive to the choice of 
the velocity/position scaling $S_{\rm G}$ which is taken here
to be equal to $S_{\rm dist}$ as discussed in previous sections.
Indeed, the value of $S_{\rm G}$ influences the
properties of the local curvature of the distribution function
(or the matrix of its second derivatives), so we expect the low-$f$
and particularly the high-$f$ regimes to be affected by the chosen
$S_{\rm G}$ (Fig.~\ref{figure3}).


\subsection{On the EnBiD position-velocity scaling}
\label{sec:posvelcor2}

\begin{figure}
\includegraphics[width=8.5cm,height=12.75cm]{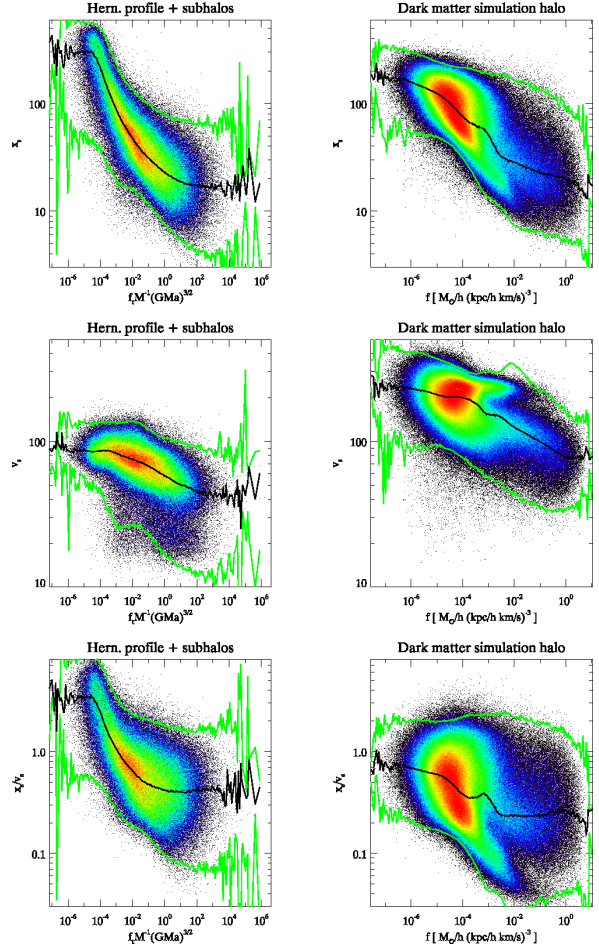}
\caption[]{Local position-velocity scaling given
by the EnBiD algorithm as a function of phase-space
  density. The {\sl left and right columns } correspond
to the composite Hernquist profile of
  \S~\ref{sec:hernquist_composite} and the simulated halo
of \S~\ref{sec:Nbodyhalo}, respectively.
{\sl From top to bottom}, the position subspace scaling, $s_x$,
the velocity subspace scaling, $s_v$, and the ratio
$S_{\rm L}=s_x/s_v$.  { \sl In  the left panels}, the phase-space density
is the theoretical one. {\sl In  the right panels}, it is measured
in the sample using SPH-AM with 40 neighbours.}
\label{figure2met}
\end{figure}

In the EnBiD algorithm, one gets for each particle 
a hyperbox of size $2 s_x$ in position subspace and of size
$2 s_v$ in velocity subspace. Figure~\ref{figure2met} shows the
measured value of $s_x$, $s_v$ and $S_{\rm L}=s_x/s_v$ 
 as functions of $f$, for our composite Hernquist profile (left
columns) and our $N$-body halo (right columns).
The global behaviour of the quantities $s_x(f)$ and $s_v(f)$ as 
decreasing functions of $f$ (the median curves in the  four upper
panels of Figure~\ref{figure2met}) is
expected, as increasing phase-space density corresponds to a smaller size for
the local hypercube. Due to the more concentrated extension of substructures
in position rather than in velocity space, the corresponding decrease is
more significant for $s_x(f)$ --by an order of magnitude-- than for
$s_v(f)$ --by about a factor 2. As a result, the ratio $S_{\rm L}=s_x/s_v$
changes from about $1.3-0.7$ in the low-$f$ regime to $0.4-0.3$ 
in the high-$f$ regime (see the median curve in two lower
panels of Figure~\ref{figure2met}).

When examining in more details the scatter plots on  the
right panels of Figure~\ref{figure2met}, we note a bimodal structure:
the cloud of points splits into two fingers at high $f$. The
shorter and denser finger corresponds to the main part of the halo,
while the other corresponds to the contribution of substructures,
which are more concentrated in phase-space than the central part of the halo.
This last statement can be easily checked by
looking at the upper right panel of Figure~\ref{figure4str}.
The main part of the halo
is globally relaxed so its concentration in velocity 
space does not depend significantly on the value of $f$
(the upper horizontal finger in the middle right panel of Figure~\ref{figure2met}),
while its position density behaves approximately like a power-law
(lower roughly straight and diagonal finger in the upper right panel of
Figure~\ref{figure2met}).
On the other hand, substructures are tidally disrupted
and lose particles while they spiral into the halo: they represent
a population of objects at various dynamical states,
different from the dynamical state of the main part of the halo. 
This explains the bimodality
observed in right panels of Figure~\ref{figure2met}.
It would however go beyond the scope of this paper
to fully explain 
the details of this bimodality.
 It 
is indeed difficult to disentangle the effect of the  local change of 
substructure phase-space/velocity subspace/position subspace
profile due to tidal deformation,
in particular to a mass loss, from the
statistical averaging carried over the population of all the substructures.

Note that, even though the prescription used to create the  
Hernquist composite profile
is dynamically unrealistic, there still 
should be a bimodal effect in this experiment, because
the substructures present a population
of objects at various ``dynamical states'', different
from the main component, owing to the fact that they
are less massive. 
However, in addition to having unrealistic individual
profiles, the contribution of substructures was purposely exaggerated: they
are more massive than those of the simulated halo. Consequently,
the bimodality is much less obvious on the left panels of
Figure~\ref{figure2met} than on the right panels. Of course, this difference
only partly accounts for this finding, as tidal stripping
changes the individual profiles of subhaloes \cite{Stoehr2006} and also
produces tidal tails that contribute in a non trivial way to
filamentation of phase-space. 

The bimodal nature of the distribution of the ratio $s_x/s_v$
has a dramatic impact on methods which rely on a global scaling between
positions and velocities prior to the measurement of the distribution
function. Furthermore, apart from that problem and the
large scatter of this ratio (of about one order
of magnitude), its median
value changes with $f$, as mentioned earlier. 
Note however the plateau reached
at high values of $f$, a regime dominated by substructures,
where $s_x/s_v \simeq 0.3$.
The global average of the ratio is equal to 0.7 and 0.5
for the Hernquist composite profile and the simulated
halo, respectively, while $S_{\rm dist}=0.4$  and 0.38.
It is important to notice that the values
of $S_{\rm dist}$ are thus close to the high-$f$
plateau, showing that high values of $f$ are expected
to be calculated with nearly optimal scaling parameter
using our peak matching of the distance distribution.

\section{Revisiting a proxy to phase-space estimation}
\label{sec:secQ}

\begin{figure}
\includegraphics[width=8.5cm,height=4.25cm]{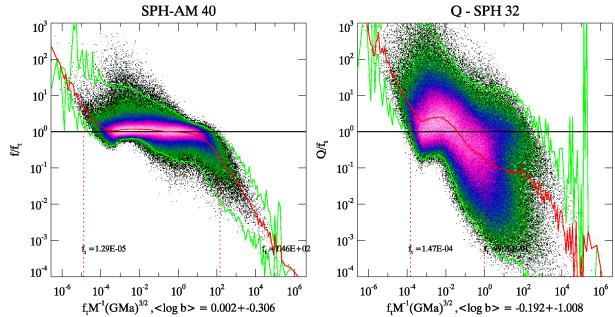}
\caption{True phase-space density estimator ({\sl left panel}) versus the
  proxy $Q=\rho/\sigma^3$ in our Hernquist composite profile. 
The ratio $f/f_t$ is shown as a function of $f_t$, where $f$ and $f_t$ are the measured and the exact
  phase-space densities, respectively. To generate the {\sl left panel},
  we used SPH-AM with 40 neighbours. To measure the function $Q({\bf x})$
  {\sl on the right panel}, we use a standard SPH interpolation 
in position space with 32 neighbours to estimate locally $\rho({\bf x})$ 
and measure the  velocity dispersion $\sigma^2({\bf x})$ with the same 
position space kernel.}
\label{figureQ}
\end{figure}

\begin{figure*}
\includegraphics[width=14cm,height=21cm]{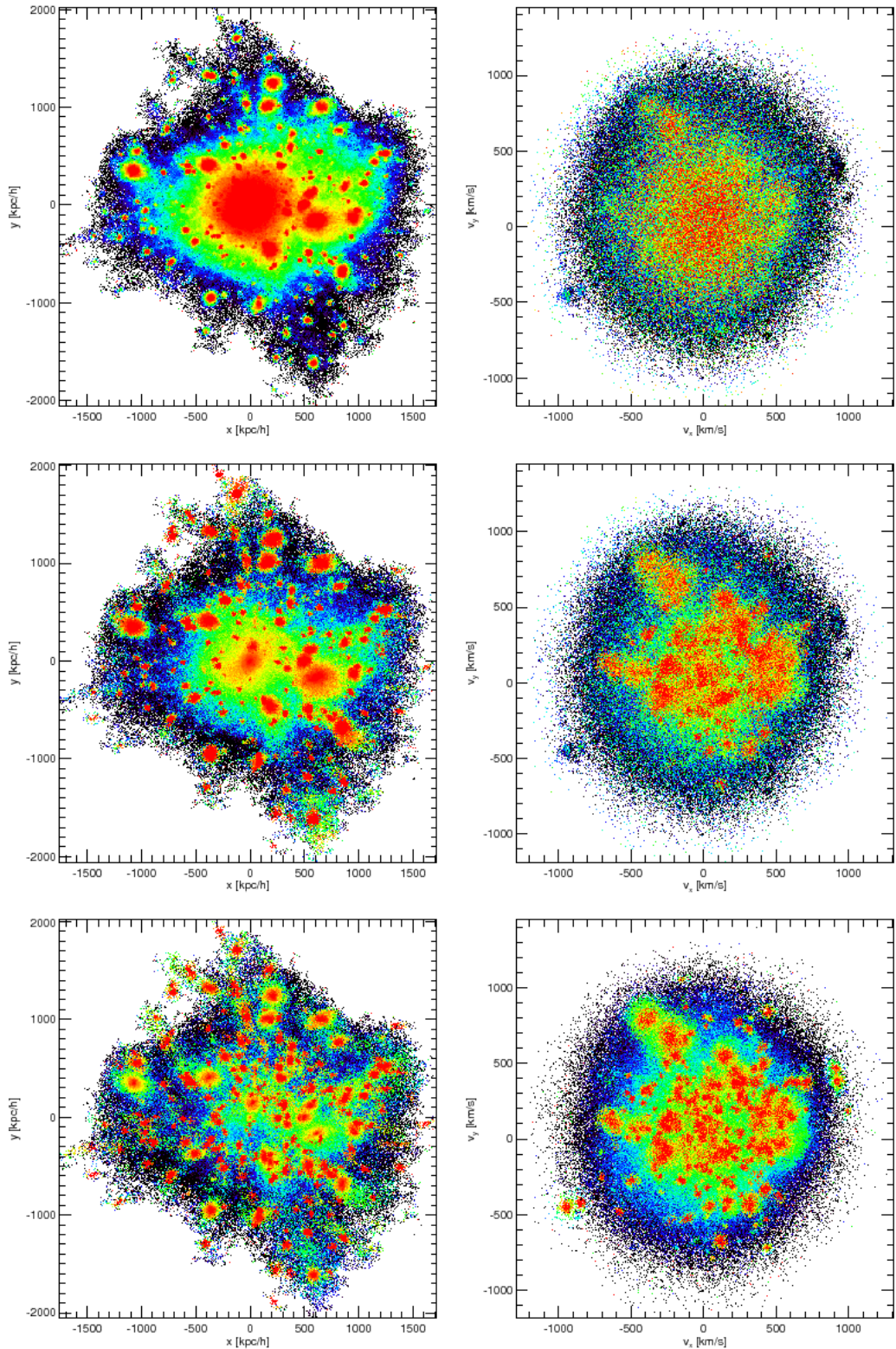}
\caption{Appearance of the CDM $N$-body haloes in position space ({\sl left
 panels}) and in velocity space ({\sl right panels}), 
with  different colour  codings. The pictures are computed in 3
steps as follows: (i) division of space into three dimensional equally spaced
grid with $N=400$ divisions across each $x,y,z$ axes, (ii)
calculation of the mean density ($\rho$,$f$,$Q$) of all particles 
inside each cell and (iii) projection of this density on the $x-y$ plane by taking
in each $z$ column the cell with the highest density.  Only
$40\%$ of  the central cells along the  $z$ axis are used for the last
step. The first, second and third rows correspond respectively to a 
colour coding with the projected density $\rho$, with
the parameter $Q=\rho/\sigma^3$ and with phase-space density $f$.   
To enhance the contrasts, the equalisation of the histograms of the
logarithm of $\rho$, $Q$ and $f$ was implemented. 
}
\label{figure6str}
\end{figure*}
\begin{figure*}
\includegraphics[width=14cm,height=21cm]{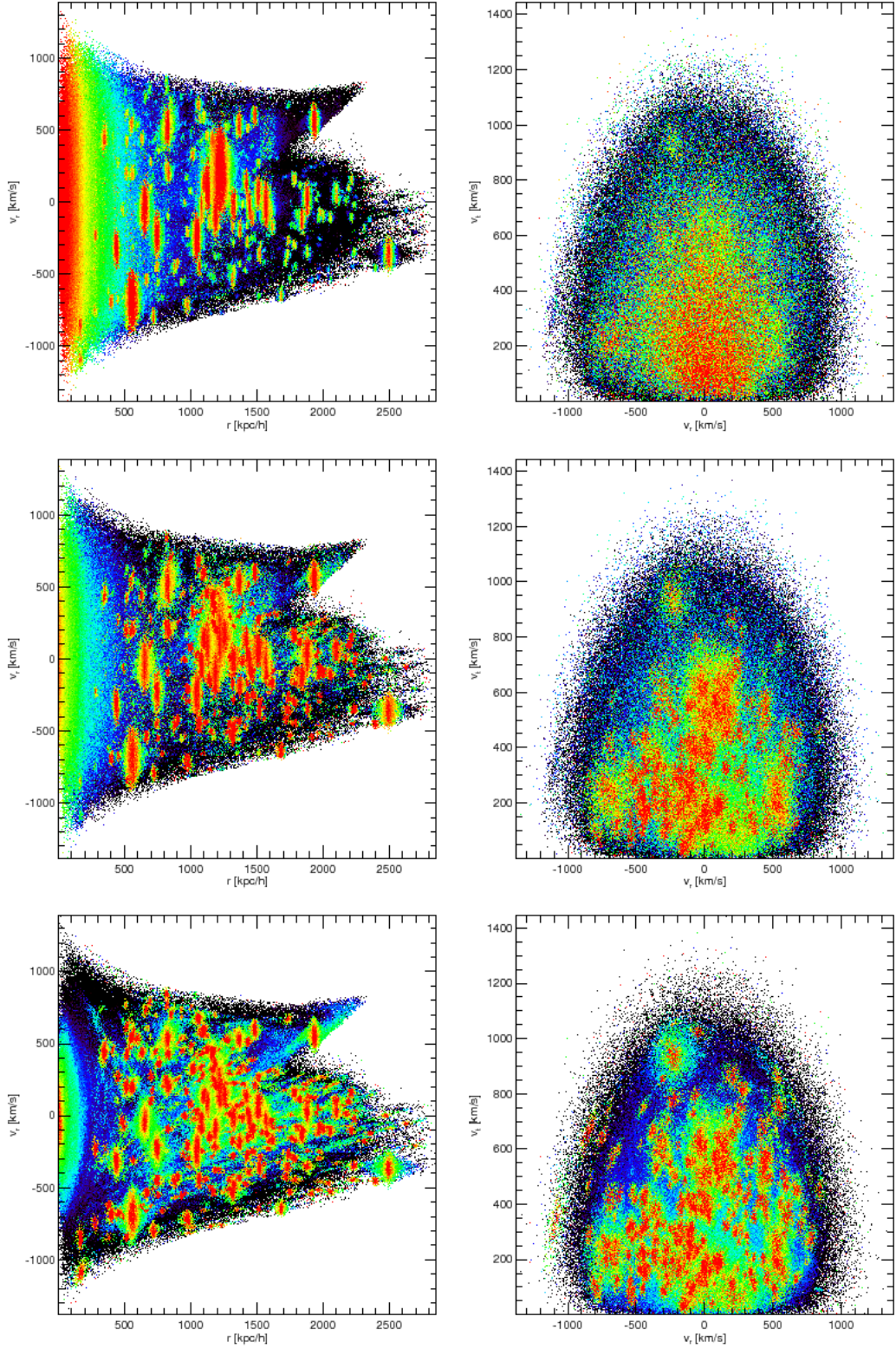}
\caption{Appearance of the CDM $N$-body haloes in radius/radial
 velocity space  ({\sl left
 panels}) and in radial/tangential velocity
 space  ({\sl right
 panels}), using the same colour coding rules as in
 Figure~\ref{figure6str}, namely using $\rho$, $Q$ and
 $f$ for the first, second and third rows, respectively.
}
\label{figure6strbis}
\end{figure*}
\begin{figure*}
\includegraphics[width=18cm,height=12cm]{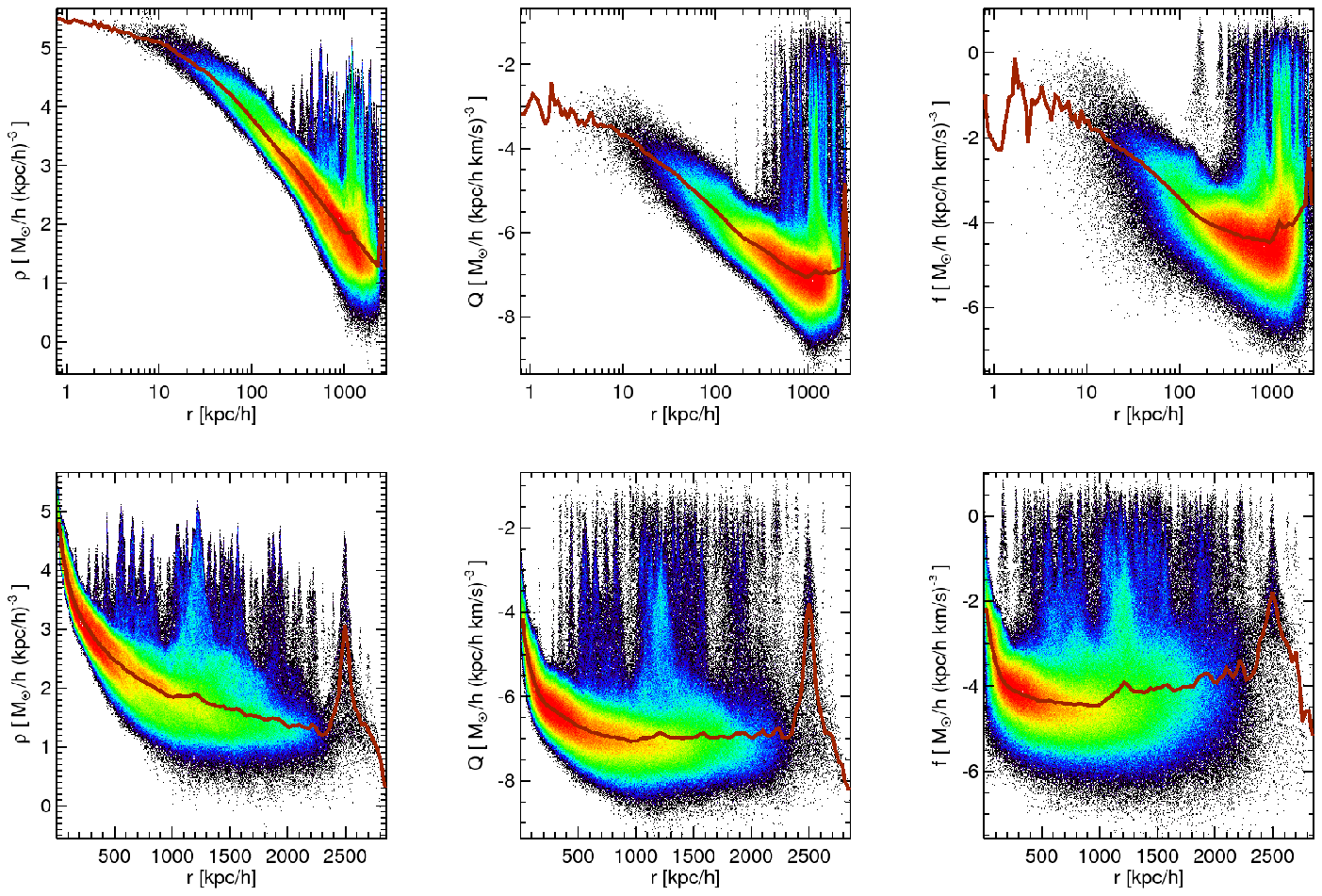}
\caption{Measured densities, {\sl from left to right}, $\rho$, $Q$ and $f$, as functions of
  distance $r$ from the halo centre in our CDM $N$-body halo.
For the
  {\sl top and bottom rows}, the density is represented as a
  function of $\log r$ and $r$, respectively, while the  thick line is calculated by taking local median.}
\label{figure4str}
\end{figure*}

Prior to the existence of 6D phase-space density
estimators, an approximation
of the phase-space density was proposed, which involves only
the measurement of quantities in position space rather than in
full 6D phase-space (see for instance, Taylor \& Navarro, 2001):
\begin{eqnarray} 
Q({\bf x}) & = & \rho({\bf x})/\sigma^3({\bf x}) \label{Qi}, \\
    & = & {3}^{3/2}\frac{\left[
    {\int\fxv {\rm d}^3v} \right]^{5/2}}{\left[ \int{v^2 \fxv {\rm d}^3v} \right]^{3/2}},
\label{eq:qtof}
\end{eqnarray}
where $\rho({\bf x})$ is the  local projected density and 
$\sigma^3({\bf x})$ is the local one dimensional velocity dispersion 
defined as $\sigma = \sqrt{(\sigma_x^2+\sigma_y^2+\sigma_z^2)/3}$.
The function $Q({\bf x})$ has been widely used in the literature
as a proxy of the true ``phase-space'' distribution function
\cite{Taylor2001,Rasia2004,Austin2005,Peirani2007,Diemand2006}. It is
often defined in a spherically
average way, $Q(r)=\rho(r)/\sigma^3(r)$.
For instance, Taylor \& Navarro (2001) found
that $Q(r) \propto r^{-\alpha}$ with $\alpha=1.875$, in good
agreement with the secondary infall model \cite{Bertschinger}.

To relate $Q(x)$ to the true phase-space density in a more
intuitive way than equation~(\ref{eq:qtof}), we can assume
that $\fxv$ factorises as follows:
\begin{equation}
  \fxv   =  \frac{\rho({\bf x})}{(2\pi)^{3/2} \sigma^3({\bf x})}
  \exp\left\{ -\frac{  [{\bf v}-{\bf v}_0({\bf x})]^2}{2
  \sigma^2({\bf x})} \right\},
 \label{eq:fap}
\end{equation}
where proper normalisations were set up directly. Hence,
\begin{equation}
 \fxv  =   Q({\bf x})
  \frac{1}{(2\pi)^{3/2}} \exp\left\{ -\frac{  [{\bf v}-{\bf v}_0({\bf x})]^2}{2  \sigma^2({\bf x})} \right\}.
\end{equation}
In particular, $Q({\bf x})=f[{\bf x},{\bf v}_0({\bf x})](2\pi)^{3/2}$.
We see that, within a normalisation factor, 
function $Q({\bf x})$ is representative of the true phase-space
distribution function where it
matters, i.e. in the neighbourhood of local maxima in velocity space.
Of course, this argument is valid only if at fixed ${\bf x}={\bf x}_0$,
the function $g({\bf v})=f({\bf x}_0,{\bf v})$ presents only one local maximum in
${\bf v}$ space. If this condition is verified, we could
expect the function $Q({\bf x})$ to represent a fair estimate of
the true phase-space distribution function near the local maxima in
phase-space, which correspond to substructures. However this is not
strictly
true since substructures a embedded in the background of the main
component
of the halo: $\sigma({\bf x})$ is not the local velocity dispersion of
the substructure but rather the local velocity dispersion 
of the diffuse component, which is much
larger.
We therefore expect $Q({\bf x})$ to {\em underestimate}
the true distribution function in substructures, corresponding to  the high
$f$ regime, which is dominated by these clumps. On the other hand, when
considering the main component of the halo, which dominates
the low $f$ regime, we expect $Q({\bf x})$ to
{\em overestimate} the true distribution function, as,  $f({\bf x},{\bf
  v}) < f({\bf x},{\bf v}_0)$ for ${\bf v}\neq {\bf v}_0$ in
equation~(\ref{eq:fap}). These arguments rely on the very simple
modeling given by equation~(\ref{eq:fap}), but they are confirmed
by Figure~\ref{figureQ}, which compares the measured function $Q({\bf x})$ to the
exact solution for the Hernquist composite profile studied
in \S~\ref{sec:hernquist_composite}. Similar trends are observed
for the simulated halo, not shown here.

%
Clearly, the function $Q$ corresponds to a serious shortcoming
 when compared
to the realistic phase-space estimators studied
in this paper. However it seems to capture the
main features of the distribution function, as illustrated by Figures~\ref{figure6str}
and \ref{figure6strbis}. 
These figures compare, in various subspaces, the structures obtained
when colour is coded by projected density $\rho({\bf x})$, 
using the parameter $Q({\bf x})$, and by phase-space density. They are
supplemented  with Figure~\ref{figure4str}, which shows $\rho$, $Q$ and
$f$ as functions of distance $r$ from the halo centre.  Note
interestingly that, both for
the 6D estimator and its proxy $Q$, 
the maximum value of phase-space density in substructures seems to be
approximately the same for all the substructures (and larger
than at the centre of the halo). This property 
is quite useful as it makes substructure
detection quite easy with simple friend-of-friend algorithm,
as proposed by Diemand et al. (2006). These authors do not use the true
phase-space distribution function but the  function $Q$ to carry the
detection.

While pure projected
density codes provide much less information than phase-space density ones, the 
$Q({\bf x})$ function  seems
to capture the most important features of phase-space, and in particular
subhaloes.
However, the true phase-space density provides additional crucial information,
in particular subtle phase-space structures such as the fine filaments
observed in the $(r,v_r)$ diagram, some of which being at the origin of 
 caustics, others corresponding to tidal tails. 
In a forthcoming work, we shall discuss  the detection and analysis
of substructures in phase-space. We shall see that 
analysis of substructures in phase-space
can be used to infer powerful properties on the dynamical
history of dark matter haloes.

\section{Summary}
\label{sec:conc}
We devoted this paper to the study of six-dimensional 
phase-space density estimators in $N$-body samples.
We considered several methods used in the literature to estimate phase-space density 
that differ from
each other (i)
in the way the  tessellation of space is performed, and (ii)
 in the way local interpolation is performed. 
Concerning point (i),
we consider two kind of tessellations: the Delaunay
tessellation (DTFE) proposed by Arad et al. (2004) through the
SHESHDEL algorithm
and the hierarchical decomposition of phase-space using
binary tree technique as proposed by Ascasibar \& Binney (2004)
through the FiEstAS algorithm, 
later improved by Sharma \& Steinmetz (2006)
with the EnBiD implementation. 
In the (ii) class, we consider two
ways of estimating the phase-space density for the DTFE method, 
one based on the direct estimation
of the local Delaunay cells volumes, and a more isotropic, smoother version
of it. For the binary tree method, we consider the hypercubical cell
smoothing proposed in FiEstAS and the  standard SPH smoothing
(but in 6D instead of 3D) using an Epanechikov kernel as
advocated by S06. We also test an anisotropic SPH method (ASPH).

In all these methods, a crucial problem is to set properly the 
local metric frame to relate position and velocities, which basically sets a scaling factor 
between the position and the  velocity subspace. While
the binary tree methods can be optimised locally both through their refinement
and through the definition of such a system of coordinates --- using a
Shannon entropy criterion, as advocated by S06 and implemented
through the  EnBiD algorithm, a global
metric must also  be defined for the DTFE method, prior to the construction
of the tessellation network. 

In order to automatically specify a global metric, we presented
two methods which yield similar results. The first one involves
simple dynamical arguments based on the  properties  of  NFW profiles. 
The second method involves 
measuring the nearest neighbour distance distributions in position
and in velocity subspace and finding the scaling factor
between position and velocities for which the positions
of the peak of these two distributions match each other. 

To summarise, we tested the following implementations:
\begin{enumerate}
\item the DTFE algorithm and a smoother, more isotropic version of it. Both
of them require
the definition of a global metric;
\item the FiEstAS binary tree method with EnBiD improvement;
\item SPH methods (a) with EnBiD improvement and (b) without it. 
We denote case (a) by SPH-AM, i.e. SPH with adaptive metric, in
opposition to case (b) that we simply denote by SPH; which requires a global metric setting;
\item Adaptive SPH methods with adaptive metric (ASPH-AM).
\end{enumerate}

To test the various algorithm in details, we used three
halo models:
\begin{description}
\item[(a)] A Hernquist isotropic profile with $5\times 10^5$ particles.
In that case, analytical estimates are available for the phase-space
distribution function.
\item[(b)] A composite Hernquist halo, built from a main
  component with $2.5\times 10^5$ particles, and a set of substructures
  amounting to $2.5\times 10^5$ particles. In that 
  more realistic case, there is also an exact expression
for the phase-space distribution function.
\item[(c)] A $N$-body halo with 1.8 millions particles 
extracted from a standard CDM simulation.
\end{description}

The main results of our analyses are the following:
\begin{itemize}
\item Because they are  local and adaptive, the  SPH-AM methods
provide the best estimators for the phase-space density, when using a
moderate number
of neighbours, ranging from 10 to 40 in order to perform the interpolation.
\item While DTFE estimators are in principle better than
SPH estimators when one measures Eulerian quantities (not centred
on the particle positions), they generally perform poorly in phase-space 
because they rely on a global metric setup. A dynamically consistent
measurement of the phase-space distribution function requires that
the scaling between positions and velocity be locally adaptive.
The best compromise, without supplementary assumption on the dynamical
history of the system, is achieved by enforcing local isotropy in
phase-space: this is achieved in practice by the Shannon
entropy criterion used in EnBiD. Note finally a last weakness of
DTFE methods: they are extremely costly from a computational point of
view compared to SPH-AM, both in  terms of computer time and memory.
\item While the optimal number of neighbours should typically be around
200 as suggested by DTFE, we find, using also DTFE, that it should
be around a few tens near high density peaks, which justifies the low
value we suggest to use  for the  SPH-AM method: such a number increases noise to signal
ratio, but allows us to probe better high phase-space density peaks.
\item By analysing the properties of the local metric
proposed by EnBiD, we find that the distance distribution matching method
provides a global scaling between positions and velocities
which probes well the high density regions
of phase-space, which are dominated by substructures. We also
find that the actual ``optimal'' local scaling presents
a bimodal distribution, made  from the contribution
of the main component of the halo, roughly in equilibrium, 
and the contribution of the substructures, which
are tidally disrupted while they spiral in  within the halo. 
This bimodality and the corresponding large scatter of about
one order of magnitude on the local scaling parameter
between positions and velocities has a dramatic
impact on the performance on methods relying on global metric
setting, such as DTFE.
\item the ASPH-AM methods do not bring much improvement over
the SPH-AM implementation. They can potentially improve
phase-space estimation in high density regions, but at
the cost of a slight systematic overestimation bias
in the moderate density regime.\footnote{Despite the fact that we
used the bias correction advocated by S06.}
\end{itemize}
Note that most of our estimators are Lagrangian in nature,{\sl i.e.} they
estimate phase-space density at particles positions: in that
sense they favour the SPH approach  relatively to the DTFE
approach. One has to keep in mind that DTFE tessellates accurately all
space, while the SPH smoothing becomes increasingly suboptimal while
departing from the particles. In particular, we found
that an Eulerian quantity such as $v(f)$ was still best measured
by DTFE, despite the problem of the suboptimal  position/velocity
scaling. 

Alternative routes to phase-space density estimation could involve
using the angle-action canonical variables which match the closest 
spherical fit to a given halo. Indeed, the topology of the underlying tori 
would provide a natural setting in which to coarse grain the distribution.

\section*{Acknowledgments}
We thank I. Arad for parts of the used code and many important comments. 
We thank S. White and V. Springel for useful discussions. Part of this 
work was completed during a visit of M.M. and S.C. at MPA/Garching. 
This work was completed within the framework of the HORIZON project 
({\tt www.projet-horizon.fr}). M. Maciejewski was funded by the 
European program $n^o$ MEST-CT-2004-504604.

\end{document}